# An AI-Driven Assistant for Human-Centric Building Digital Twin Systems: Design, Implementation, and Comparative User Evaluation

Leonardo Messi[a, *], Erika Anneli Pärn[a,b], Juan Diego Castaño Molina[a], Oleksandr Anyshchenko[a], Dijar Selami[a], Peihang Luo[b], and Borja García de Soto[a]

[a]S.M.A.R.T. Construction Research Group, Division of Engineering, New York University Abu Dhabi (NYUAD), United Arab Emirates (UAE)

[b]Construction Information Technology Laboratory, Department of Civil Engineering, University of Cambridge, Cambridge, United Kingdom (UK)

[*]Corresponding author: Leonardo Messi, lm4720@nyu.edu

## Abstract

Smart buildings generate streams of operational data, yet this richness rarely translates into accessible, decision-ready information for stakeholders who inhabit and manage them. Current Building Digital Twin Systems (BDTS) often remain oriented toward technical specialists and rely on static, one-size-fits-all interfaces. This paper presents the design, implementation, and comparative user evaluation of an AI-driven conversational assistant integrated into a human-centric BDTS for Indoor Environmental Quality monitoring, combined with adaptive information presentation and auto-suggest functionality. The prototype was deployed at New York University Abu Dhabi. A controlled, two-arm, between-subjects, quasi-experimental study with wave-based, non-randomized assignment compared the Baseline interface with the Experimental interface. Results showed significant reductions in task-completion time and perceived workload, while accuracy and usability improved but did not meet the adjusted significance criterion. The study contributes an implemented and empirically evaluated architecture that separates generative interaction from deterministic environmental assessment and coordinates conversational responses with adaptive visual presentation.

## 1. Introduction

The digitalization of the built environment has accelerated substantially over the past decade (Papadonikolaki et al., 2022). Smart buildings, infrastructure, and industrial facilities are increasingly equipped with dense sensor networks, Internet of Things (IoT) devices, and sophisticated monitoring and control systems that collectively produce large volumes of operational data (Chaudhari et al., 2024; Lavrinovica et al., 2024; Pluth et al., 2025). From sub-minute energy consumption logs to real-time occupancy tracking, from particulate matter concentrations to HVAC setpoint histories, modern buildings and facilities possess a remarkable capacity for self-monitoring (Chaudhari et al., 2024; Pluth et al., 2025). However, the growing availability of operational data does not automatically make such information accessible, understandable, or actionable for the people who occupy and manage these environments (Ceccarini et al., 2022; Pluth et al., 2025).

This disconnect constitutes a human-interface gap in smart building management. The problem arises less from a scarcity of data or computational infrastructure than from a mismatch between expert-oriented Building Digital Twin System (BDTS) interfaces and the diverse professional, cognitive, and contextual needs of their users (Osei-Asiamah et al., 2025). Facility managers may struggle to extract actionable insights from complex operational data, while occupants may lack intuitive means of understanding the indoor environmental conditions that affect their comfort and well-being. Consequently, the potential value of data-rich buildings may remain underexploited because their information is difficult to retrieve, interpret, and translate into appropriate responses (Lou et al., 2025).

This paper addresses this gap through the design, implementation, and comparative user evaluation of an AI-driven conversational assistant integrated into a human-centric BDTS. The proposed assistant acts as an interaction layer between users and the underlying building information environment, supporting natural-language information retrieval, contextual explanation, and grounded visual outputs. The Experimental interface further integrates adaptive information presentation and auto-suggest functionality, with the aim of reducing reliance on complex dashboard hierarchies and specialist querying knowledge. The system was developed according to three literature-informed principles: stakeholder- and role-aware information delivery, cognitive-load-aware information prioritization, and interpretable IEQ support.

The prototype was deployed at the New York University Abu Dhabi (NYUAD) campus, with the SMART Construction Research Group laboratory serving as the principal testbed. Its comparative evaluation followed a two-arm, between-subjects, post-test-only quasi-experimental design. The conventional BDTS interface was compared with an Experimental interface integrating an AI-driven assistant, hereafter referred to as the Copilot, adaptive presentation, and auto-suggest functionality. Evaluation combined objective task accuracy and time-on-task data with standardized measures of perceived workload and usability, complementary user-experience items, feature-engagement data, and qualitative feedback.

This paper makes three principal contributions. First, it derives and operationalizes literature-informed design principles for role-aware, cognitive-load-aware, and interpretable interaction with Building Digital Twin Systems. Second, it develops an assistant-centered architecture that integrates natural-language access to heterogeneous BDTS information with adaptive visual presentation, while separating generative language functions from deterministic environmental assessment and citation grounding. Third, it provides comparative empirical evidence on the effects of this integrated interface on task effectiveness, interaction efficiency, perceived workload, and usability relative to a conventional BDTS interface. The individual components of the architecture, namely conversational access, retrieval-augmented generation, and adaptive presentation, build on established approaches; the contribution lies in their coordination within a BDTS, in particular the separation of deterministic environmental assessment from language generation, and in the comparative evaluation of the resulting integrated interface rather than of its individual components.

The remainder of this paper is organized as follows. Section 2 reviews the relevant scientific background and formalizes the research questions. Section 3 presents the methodology, including the design principles and system architecture. Section 4 describes the technical implementation and the NYUAD testbed. Section 5 describes the comparative validation protocol for the Baseline and Experimental UIs. The results and their implications are discussed in Section 6. Finally, Section 7 summarizes the study’s contributions, limitations, and directions for future research.

# 2. Scientific Background

## 2.1. Building Digital Twin Systems

A Digital Twin (DT) is commonly described as a digital representation of a physical object, process, or system that is maintained using information from its physical counterpart (Bae et al., 2025). Conventional accounts characterize a DT through three related elements: a physical entity in physical space, a corresponding digital entity in digital space, and a persistent bidirectional data connection between them that enables real-time synchronization between the two (Bae et al., 2025; Boje et al., 2020; Liu et al., 2024; Mousavi et al., 2024; Tahmasebinia et al., 2023). Unlike static digital models or one-way digital shadows, mature DT implementations support continuous or near-real-time data exchange between the physical and virtual environments. Where control and actuation mechanisms are available, this exchange may also enable feedback from the virtual model to the physical system (Bae et al., 2025; Barham et al., 2025; Mousavi et al., 2024).

The conceptual precursors of DT technology are often associated with NASA's use of mirrored systems during the Apollo program, including the Apollo 13 mission, to simulate and address failures in a remote spacecraft (Bae et al., 2025; Omrany et al., 2023; Su et al., 2025; Tahmasebinia et al., 2023; Zhao et al., 2022). The contemporary DT concept is more directly traceable to Michael Grieves's work in Product Lifecycle Management. Grieves publicly presented an unnamed physical-virtual product model in 2002, commonly identified by the title "Conceptual Ideal for PLM" (Liu et al., 2024; Mousavi et al., 2024; Omrany et al., 2023; Tahmasebinia et al., 2023). Since then, the technology has evolved from a theoretical concept into a sophisticated framework enabled by Industry 4.0 technologies, such as the Internet of Things (IoT), Big Data, and Artificial Intelligence (AI) (Ma et al., 2025; Tahmasebinia et al., 2023; Yitmen et al., 2025).

Beyond the conventional component-based definition of an individual DT, Davila Delgado and Oyedele (2021) distinguish between the structural and functional descriptions of DT systems. Structural descriptions comprise conceptual models, system architectures, and data models, whereas functional descriptions explain the processes and communication mechanisms through which physical and digital components interact. Pärn et al. (2025) extend this systems-oriented perspective through the concept of Twin Systems, understood as interconnected networks of physical and digital products and processes that evolve together across the built-asset lifecycle. This framing supports the use of the term Building Digital Twin System (BDTS) in the present study, as the proposed contribution concerns not only the virtual representation of a building but also the sensing, data, analytical, and interaction services through which it operates. Furthermore, Fang et al. (2025) argue that built-asset DTs should progress beyond predominantly geometric 3D representations toward behavioral twins capable of dynamically representing asset states and behavior throughout the lifecycle. In a similar vein, Twin Systems by Pärn et al. (2025) defines the digital twin in terms of design intent information, status information, and software functions for comparison, performance assessment, prediction, optimization, and operational communication.

Owing to its versatility, DT technology has been applied across a wide range of domains. In aerospace and aeronautics, DTs have been used to support structural life prediction, predictive maintenance, and mission flight simulation (Bae et al., 2025; Boje et al., 2020; Mousavi et al., 2024; Yang et al., 2024). In manufacturing, they contribute to the development of smart factories by enabling the optimization of production workflows and the real-time diagnosis of equipment faults (Bae et al., 2025; Boje et al., 2020; Liu et al., 2024; Su et al., 2025; Tahmasebinia et al., 2023; Yang et al., 2024). DT applications have also expanded into healthcare, where they support precise patient monitoring and personalized treatment planning (Bae et al., 2025; Liu et al., 2024; Tahmasebinia et al., 2023; Venkateswarlu and Sathiyamoorthy, 2025; Yang et al., 2024). Similarly, in the energy and urban planning sectors, DTs are used to optimize power generation and support large-scale infrastructure initiatives, such as the Virtual Singapore project (Bae et al., 2025; Mousavi et al., 2024; Tahmasebinia et al., 2023; Yang et al., 2024). Within the Architectural, Engineering, Construction and Operations (AECO) industries, BDTSs offer a lifecycle-oriented framework that can support activities from design and construction monitoring to operational facility management (Davila Delgado and Oyedele, 2021; Long et al., 2024; Mousavi et al., 2024; Omrany et al., 2023; Zhao et al., 2022). Current applications encompass project planning, safety management, structural health monitoring, quality control, energy performance, and occupant comfort (Long et al., 2024; Mousavi et al., 2024; Omrany et al., 2023). However, Long et al. (2024) found that existing research remains concentrated primarily in the construction and Operation and Maintenance (O&M) phase, while other lifecycle stages receive comparatively limited attention. Their review also identifies stakeholder integration, technological limitations, and the absence of universal industry standards as persistent barriers to lifecycle implementation.

Beyond technical implementation, the long-term adoption of Twin Systems also depends on their capacity to generate organizational value through four principal routes: the Game Changer, Efficiency Gainer, Added Value Provider, and Platform Builder models (Pärn, 2025). These models emphasize the need to align digital services with stakeholder requirements and organizational value propositions.

A fundamental challenge in BDTSs lies not only in the volume of collected data but also in the semantic integration, reliability, and retrieval of heterogeneous information. BDTSs combine static and dynamic data, including BIM models, point clouds, environmental measurements, mechanical system states, and real-time sensor streams. Kosse et al. (2025) address this challenge through a modular Semantic Digital Twin architecture based on information containers, enabling heterogeneous information to be organized and exchanged according to changing lifecycle and use-case requirements. Data reliability is equally critical: Koo and Yoon (2024) demonstrate that the simultaneous calibration of physical and virtual sensors can improve synchronization reliability between physical and digital building systems. Beyond data integration and calibration, AI-based methods can support operational diagnostics, as illustrated by the application of autoencoders to fault detection using building automation system data (El Mokhtari and McArthur, 2024).

BDTSs are increasingly used to monitor and support the management of Indoor Environmental Quality (IEQ), which influences occupant health, productivity, and well-being (Yitmen et al., 2025). By integrating IoT sensor networks with 3D models, BDTSs can provide spatiotemporal insights into key environmental parameters, including air quality indicators such as $CO_2$ and VOCs, thermal comfort, acoustics, and lighting conditions (Zaballos et al., 2020; Zheng et al., 2025). Several recent studies illustrate the growing relevance of DT-based approaches for IEQ assessment and management. Zaballos et al. (2020) proposed a smart campus DT to calculate an IEQ index based on environmental monitoring and occupant emotions. Yitmen et al. (2025) investigated AI-driven BDTSs for enhancing IEQ and energy efficiency, emphasizing decision support for automated fault detection. Zheng et al. (2025) presented a scalable framework for assessing IEQ in office spaces, demonstrating thermal comfort visualization in 3D environments. Similarly, Opoku et al. (2024) applied a DT platform to monitor indoor conditions in Living Labs, including university library environments, while Ma et al. (2025) developed an architecture for 3D interactive operations, including AI-empowered HVAC optimization for indoor comfort. Kang and Mo (2024) proposed a digital twin framework and architecture emphasizing real-time connectivity and predictability, evaluated on the monitoring of the environmental performance of existing buildings, while ElArwady et al. (2024) integrated BIM, IoT, and machine learning within a digital twin platform for modeling indoor thermal comfort. Karatzas et al. (2024) further used a text-analytic framework to examine the integration of digital twins and machine learning, treating Indoor Environmental Quality, energy efficiency, and occupant comfort as key facets of indoor building performance. In educational settings, Günzel et al. (2026) reported that smart-controlled rooms maintained lower $CO_2$ levels and more stable temperatures, improving concentration and perceived comfort.

The management of IEQ and occupant comfort is a cross-cutting concern affecting all building stakeholders, from facility managers, who must balance operational and maintenance requirements, to occupants, who directly experience indoor conditions (Liu et al., 2025; Yitmen et al., 2025; Zaballos et al., 2020). Traditional Building Automation Systems (BAS) often rely on static, expert-oriented 2D interfaces that can contribute to information overload and provide limited support for the dynamic needs of different user groups (Ma et al., 2025).

Collectively, these limitations motivate the adoption of human-centric design and human-in-the-loop (HITL) principles, through which building stakeholders become active participants in the interpretation and use of BDTS information. The implications of these principles for adaptive interface design are examined in the following section.

## 2.2. Adaptive User Interfaces in Complex Sociotechnical Systems

Against this background, human-centric design and human-in-the-loop (HITL) principles should be operationalized as explicit information and interaction requirements within BDTSs. The interface should not merely expose available data but should organize and communicate information according to users' goals, responsibilities, expertise, and decision contexts. Čustović et al. (2025) observe that the technical development of construction DTs has progressed more rapidly than the investigation of human performance factors. Their analysis demonstrates that the information required to maintain situation awareness varies across users and operational contexts, supporting the development of context-sensitive DT interfaces that align information delivery with human decision-making needs.

The evolution of smart buildings into complex sociotechnical systems has revealed a significant gap in how building data are communicated to occupants and operational stakeholders. Occupants often remain passive recipients of environmental information about their workspaces (K. S. Lee et al., 2025). Traditional BAS interfaces typically present data through static, expert-oriented interfaces composed of 2D diagrams and extensive numerical tables, which can lead to information overload for non-technical users (Ma et al., 2025). Similar limitations have been identified in human-building interaction and construction DT research, where the interface is increasingly recognized as a critical layer for translating building data into situational awareness, occupant feedback, and decision support (Ceccarini et al., 2022; M. J. Lee et al., 2025; Soman et al., 2025; Yoon et al., 2025). This lack of transparency and the difficulty of navigating complex datasets can reduce awareness of building performance and of the environmental factors affecting personal well-being (K. S. Lee et al., 2025). Such barriers may contribute to occupant dissatisfaction; K. S. Lee et al. (2025) report that up to 81% of occupants are dissatisfied with at least one aspect of their workspace, a challenge compounded by limited access to understandable information about immediate environmental parameters such as temperature and lighting.

To bridge this information gap in smart buildings, K. S. Lee et al. (2025) proposed an occupant-centric interface framework that prioritizes observability and informed decision-making. The framework comprises three complementary interaction modes to help users retrieve information according to their needs: person-based, object-based, and recommendation-based interaction. Person-based interaction allows users to view the status of their environment in relation to psychological and physiological states, focusing on how they experience the space rather than requiring them to interpret technical setpoints. Object-based interaction enables users to retrieve real-time information by selecting positioned 3D virtual objects within the Digital Twin, such as fans or lighting fixtures, without requiring expert knowledge of the BAS hierarchy. Recommendation-based interaction leverages AI-driven predictive analytics to proactively inform users about environmental changes and, through explainable AI (XAI), provides visual and textual explanations that support trust through transparent information sharing.

For occupant-centric interfaces to effectively support information retrieval, transitions between these interaction modes should occur dynamically and automatically according to users' current context, cognitive requirements, and interaction behavior (Osei-Asiamah et al., 2025). Requiring users to manually configure the interface increases cognitive effort and may hinder efficient access to relevant information. Conversely, an adaptive interface can reduce this burden by automatically highlighting the most relevant data and interaction options at the appropriate moment. In this way, the DT evolves from a complex repository of building data into a human-centric decision-support system that enables occupants to better understand, interpret, and interact with their built environment (K. S. Lee et al., 2025).

Adaptive User Interfaces (AUIs) can be implemented through different adaptation logics and enabling technologies, including rule-based mechanisms, model-driven engineering, context-aware systems, accessibility profiles, recommendation engines, machine learning, and LLM-based agents. AI is therefore not a defining requirement of an AUI: rule-based and model-driven approaches remain effective when adaptation criteria, such as user profiles, device constraints, or accessibility requirements, can be specified explicitly and remain sufficiently stable (Zouhaier et al., 2023). Context-aware and learning-based approaches instead enable the interface to respond dynamically to user behavior, task conditions, and environmental changes (Carrera-Rivera et al., 2024).

Recent studies provide relevant methodological precedents for this transition. Wang et al. (2025) developed a context-aware adaptive information-push method that adjusts information delivery according to users' cognitive abilities, including skill level and fatigue, demonstrating reductions in task time, errors, and cognitive load. Luo et al. (2026a) developed a personalized alert generation system that tailors the generic alerts generated by the digital twins to the role and skill level of the recipients, demonstrating reductions in the information overload rate and improvements in the role-content match rate. Similarly, Muthu Bharath et al. (2026) proposed the X-DT framework, integrating role-specific adaptive dashboards with rule-based assessment to support context-aware building performance interpretation. Xie et al. (2025) further showed that alternative interaction modalities produce different cognitive-load and user-experience outcomes, emphasizing the importance of empirically evaluating the interaction mechanisms adopted by adaptive systems. In the construction domain, Wu et al. (2025) integrated LLMs, retrieval-augmented generation, hierarchical knowledge representation, and user-preference learning to support accessible information retrieval and question answering. Although these studies were not developed specifically as IEQ-oriented BDTS interfaces, they demonstrate how context modeling, interaction data, and AI-based retrieval can be combined to adapt information delivery to users and tasks.

These capabilities are particularly relevant to BDTS, where users must retrieve, interpret, and act upon heterogeneous technical and environmental data. LLM-enabled mechanisms can support natural-language interaction, infer user intent, adjust explanation depth, and generate context-sensitive recommendations. In the construction sector, Saka et al. (2024) assessed GPT models across the project lifecycle, discussing their opportunities and challenges and validating a BIM-based use case for material selection and optimization. Recent systems such as ChatTwin, Graph-DT-GPT, HighwayMAS, and TailorAlert further demonstrate the transition toward conversational, agent-based, and user-aware interaction models (Luo et al., 2026a, 2026b; Pan et al., 2026). AI-driven AUIs therefore represent a promising direction for transforming BDTS from static visualization platforms into adaptive, human-centric decision-support systems.

To clarify the positioning of the proposed contribution, Table 1 compares the proposed system with the closest related Digital Twin interaction systems across eight capabilities spanning conversational access, information grounding, access to dynamic and model-based data, interface adaptation, domain assessment, and empirical evaluation. The comparison distinguishes capabilities that were explicitly implemented and demonstrated from those that were only partially or indirectly addressed.

As summarized in Table 1, previous studies have advanced complementary components of human-centric Digital Twin interaction. Occupant-centric frameworks have explored context-dependent interface modes, while ontology- and graph-based approaches have improved access to structured Digital Twin information. More recent LLM-based systems have enabled natural-language interaction, grounded information retrieval, role-specific alerts, and multi-agent orchestration. However, among the reviewed studies, none combines natural-language access to live IEQ and BIM/model information, grounded retrieval, role-conditioned adaptive visual presentation, deterministic environmental assessment, and empirical comparison with a conventional BDTS interface. The proposed system addresses this integration gap by coordinating these capabilities within a single human-centric Building Digital Twin System.

*Table 1. Capability comparison between the proposed system and closely related Digital Twin interaction studies. ✓ = implemented; ◐ = partially addressed; — = not reported.*

| Study/ system | Natural-language interaction | RAG/ structured grounding | Live sensor information | BIM/ model querying | Adaptive visual interface | Role-conditioned output | Deterministic environmental assessment | Empirical comparison with a conventional interface |
|---|---|---|---|---|---|---|---|---|
| (K. S. Lee et al., 2025), occupant-centric DT interfaces | — | — | ◐ | ◐ | ◐ | — | — | — |
| (Yoon et al., 2025), ontology-enabled intelligent DT | ◐ | ✓ | ◐ | ◐ | — | — | — | — |
| (Muthu Bharath et al., 2026), X-DT | — | — | ◐ | — | ✓ | ✓ | ✓ | — |
| (Luo et al., 2026a), TailorAlert | ◐ | ◐ | ◐ | — | — | ✓ | — | — |
| (Luo et al., 2026b), ChatTwin | ✓ | ✓ | ✓ | ◐ | — | — | — | ◐ |
| (Pan et al., 2026), Graph-DT-GPT | ✓ | ✓ | — | ✓ | — | — | — | — |
| (Lu et al., 2026), HighwayMAS | ✓ | ✓ | ◐ | ◐ | — | — | — | ✓ |
| **Proposed AI-adaptive BDTS** | ✓ | ✓ | ✓ | ✓ | ✓ | ✓ | ✓ | ✓ |

## 2.3. Research Questions

Taken together, existing research has advanced several components required for human-centric BDTS, including lifecycle-oriented frameworks, semantic data architectures, behavioral twins, sensor reliability, AI-enabled diagnostics, human-centered information requirements, context-aware information delivery, and LLM-based retrieval. Recent studies have shown natural-language access, agent-based interaction, personalized alerts, and AI-assisted information retrieval in Digital Twin environments. However, evidence remains limited regarding systems that combine grounded conversational access with coordinated adaptation of the underlying visual interface and that empirically evaluate the resulting human-interaction benefits against a conventional BDTS interface. In addition, environmental assessments generated by LLM-enabled systems are typically dependent on generative reasoning rather than explicitly separating probabilistic language processing from deterministic threshold evaluation. Evidence from built-environment applications also indicates that LLM-generated advisory outputs should be interpreted cautiously within decision-making processes (Koohsari et al., 2026). Evidence also remains limited regarding whether an AI-adaptive interface can reduce the effort required to retrieve and interpret building information while maintaining or improving task effectiveness and usability. This gap motivates the present study, which is structured around the following research questions:

**RQ1.** What design principles can guide the development of an AI-driven AUI for human-centric BDTS?

**RQ2.** To what extent does the proposed AI-driven AUI improve task effectiveness, interaction efficiency, usability, and hedonic user experience, while reducing perceived workload, compared with the baseline BDTS UI?

**RQ3.** How does the proposed AI-driven AUI affect users' access to and retrieval of IEQ information within a BDTS?

## 3. Methodology

To address the research gap identified in Section 2, the study adopts a design-and-evaluation methodology comprising the formulation of literature-informed design principles, the development and implementation of the proposed system, and its comparative validation against a non-adaptive baseline. The design principles establish the human-centric and adaptive requirements of the system, while the subsequent methodological stages describe its architecture, technical implementation, deployment, and comparative validation.

### 3.1. Design Principles

Building on the literature synthesis and the research questions, three design principles were formulated to guide the development of an AI-driven AUI for a human-centric BDTS. Consistent with the systems-oriented perspective introduced in Section 2, the proposed BDTS integrates building and sensor data, AI-based interpretation of the current request, declared stakeholder role, and active conversational context, adaptive information presentation, and human interaction within a unified sociotechnical system.

The first design principle is stakeholder- and role-aware information delivery (DP1). The interface should support differentiated information delivery according to available user and interaction context, including stakeholder role, task requirements, and, where available, expertise or preference information. Facility managers, technical operators, researchers, and occupants may require access to the same underlying building data but at different levels of detail and through different forms of representation. The system should therefore replace a one-size-fits-all interface with differentiated interaction mechanisms for heterogeneous stakeholders (Čustović et al., 2025; K. S. Lee et al., 2025).

The second design principle is cognitive-load-aware information prioritization (DP2). Rather than exposing users directly to complex and high-density BDTS data streams, the interface should filter, prioritize, and organize information according to the current task and interaction context. This principle aims to reduce unnecessary navigation and interpretation effort while enabling users to identify relevant information more efficiently (Hamdani and Chihi, 2025; Wang et al., 2025).

The third design principle is interpretable and actionable IEQ decision support (DP3). The interface should present IEQ indicators together with their spatial and temporal context, explanatory information, and decision-relevant implications. Users should be supported not only in accessing environmental measurements but also in understanding their significance for comfort, well-being, and building operation. This principle enables both occupants and facility stakeholders to interpret indoor conditions and identify appropriate responses (K. S. Lee et al., 2025; Yitmen et al., 2025; Zaballos et al., 2020).

In the implemented prototype, these principles are operationalized through three forms of adaptation: (i) query-responsive interface adaptation, whereby the presentation changes according to the current information request; (ii) role-conditioned response adaptation, based on an explicitly declared stakeholder role; and (iii) conversational-context adaptation within the active session. Persistent modeling of user expertise, preferences, and long-term behavior was outside the scope of the present implementation.

Together, these principles provide the design-oriented response to RQ1. The implemented adaptation mechanisms are evaluated under RQ2 through measures of task effectiveness, interaction efficiency, perceived workload, usability, and hedonic user experience, while DP3 provides the methodological basis for investigating users' access to and interpretation of IEQ information under RQ3.

## 3.2. System Architecture

Building on the design principles introduced in Section 3.1, the proposed architecture addresses all three design principles. Cognitive-load-aware information prioritization (DP2) is served by scoping the data retrieved for each question and by adapting the interface that accompanies the answer; interpretable and actionable IEQ decision support (DP3) by computed threshold verdicts and traceable citations; and stakeholder- and role-aware information delivery (DP1) by a declared role that shapes how each answer is written. The Copilot mediates access to heterogeneous BDTS data sources, reducing the need for users to navigate multiple interfaces and data structures, while grounding its responses in spatial, historical, real-time, and forecast information.

Figure 1 organizes the human-centric BDTS into four interconnected layers, arranged from bottom to top: the sensing layer, the data and model services layer, the application services layer, and the user interface layer. Environmental sensors publish their readings over MQTT, and a time-series database stores them each minute. The data and model services layer holds a database storing the sensor time series, building models, and Copilot knowledge base. The building models represent spaces and elements through BIM and IFC geometry, point clouds, and 3D Gaussian Splatting (3DGS). The Copilot knowledge base holds curated knowledge cards, guideline records carrying published threshold values with their sources, and the conversation history used to resolve references across turns. Table 2 summarizes the sources consulted at query time.

Application Services fronts these sources through a single platform API (the Digital Twin Platform API of Figure 1), which serves two consumers: the web application above it and the IEQ Copilot engine alongside it. The Digital Twin Platform API supplies the measurements, the building model, and the twin assets that the interface renders, which is the pathway a user follows when exploring a space visually. The IEQ Copilot engine runs as a separate service and reaches the same sources through the DT Platform API, returning a grounded answer and a metadata event that adapts the User Interface Layer around it. Concentrating data access in one API keeps the platform usable on its own and leaves a single place where that access is authorized and extended; Section 4.2 reports the server-side implementation.

In the User Interface Layer at the top of the stack, building users reach the system through space-specific dashboards, the interactive 3D twin viewer, and the Copilot chat, delivered by one web application. Visual exploration and conversation therefore remain available side by side, and Section 4.1 describes the interface itself.

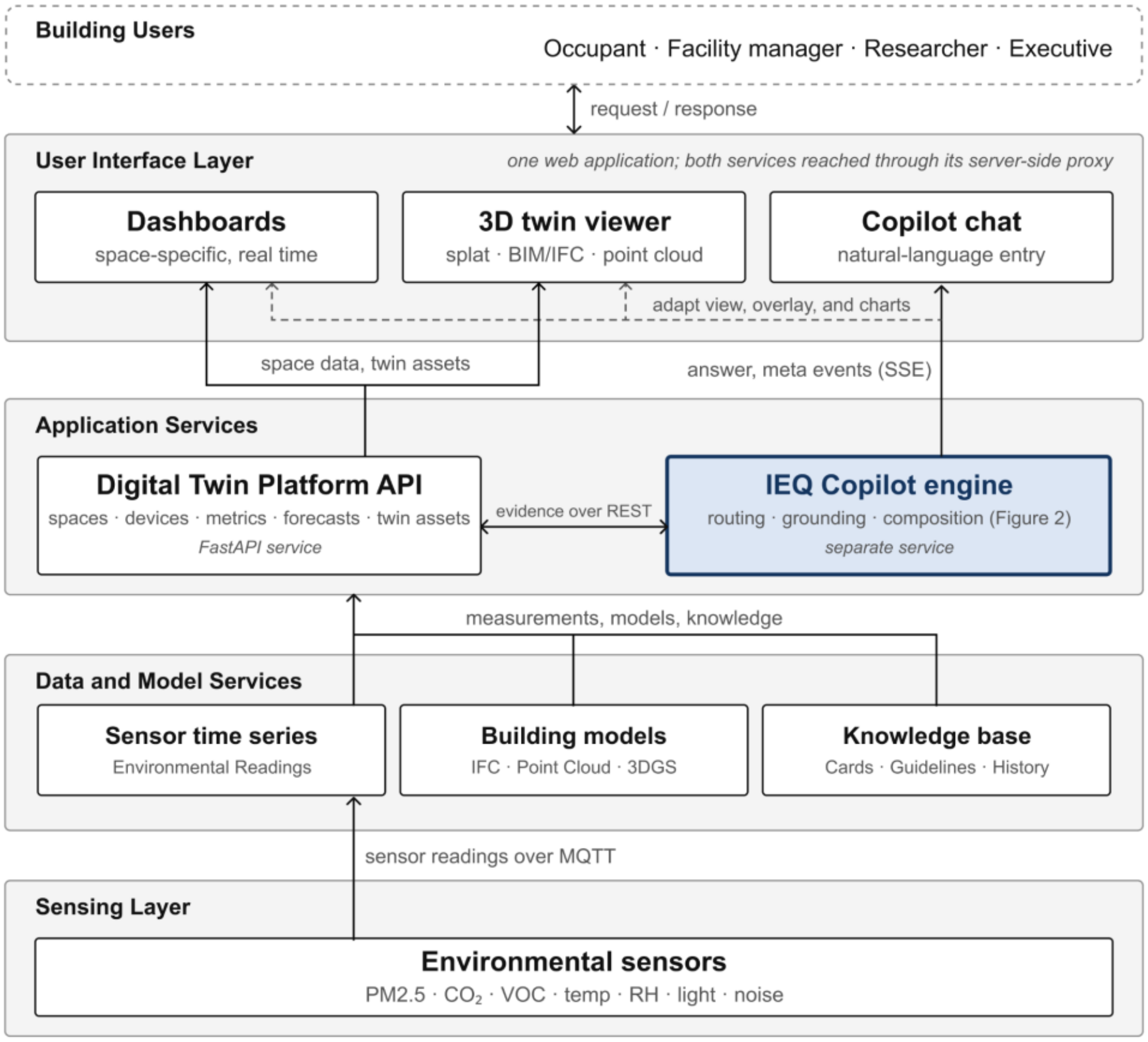

*Figure 1. High-level architecture of the human-centric BDTS. The Digital Twin services and the Copilot engine consume the data and model services as peers, so the platform remains usable without the assistant. The Copilot engine is detailed in Figure 2.*

## 3.3. IEQ Copilot engine

Figure 2 shows how the IEQ Copilot engine processes a single request. The workflow is drawn in BPMN 2.0 (Object Management Group, 2013), with one lane for each of the four internal layers: the Interface Layer, Context and Routing, Evidence Retrieval, and Grounding and Composition. A request enters through the Interface Layer, passes through the other three layers in turn, and returns to the Interface Layer for delivery. The data sources consulted along the way are listed in Table 2 together with the step at which each enters the flow. The following sections trace this path one layer at a time, using the step labels of the figure.

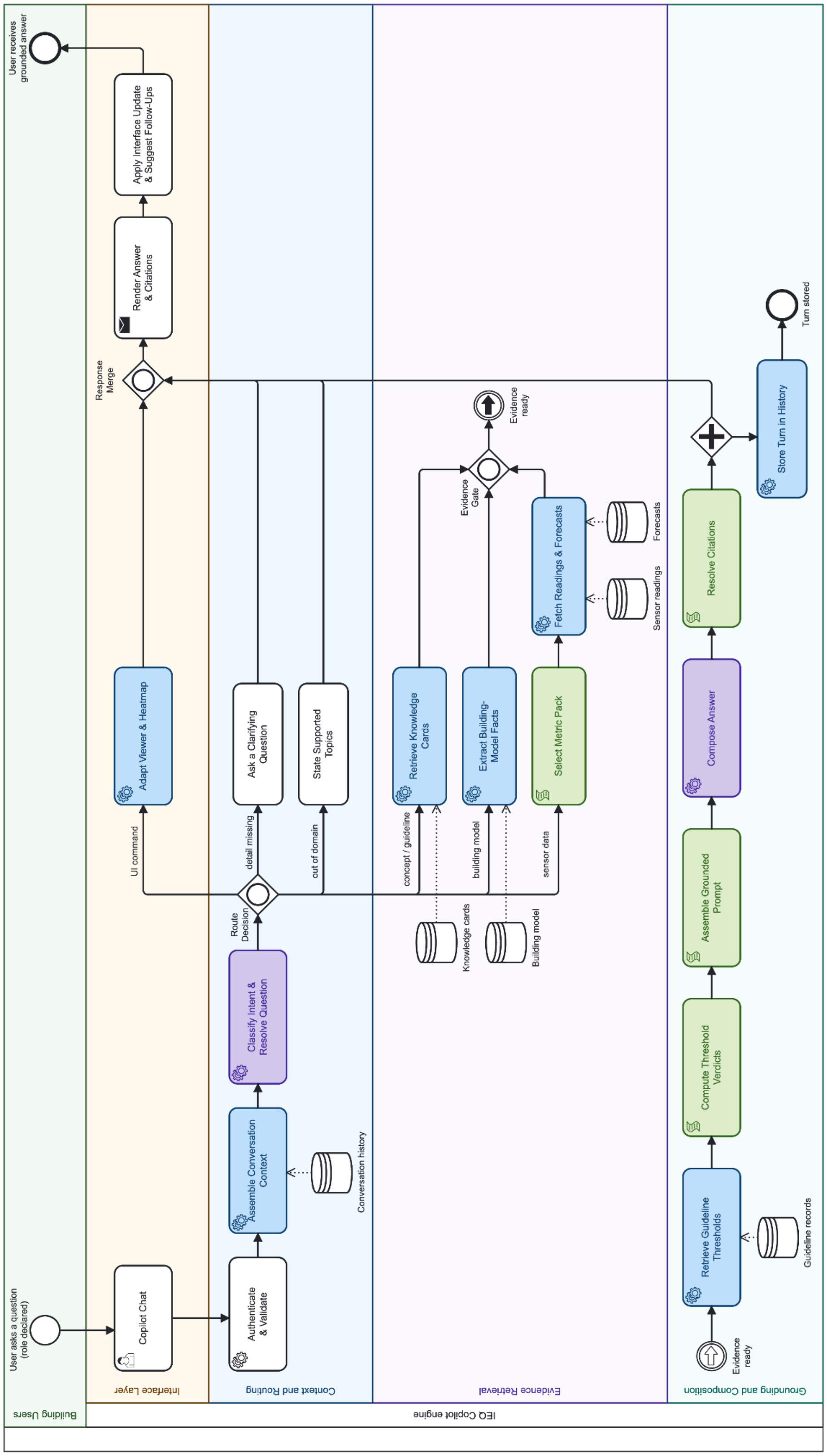

*Figure 2. Request-processing workflow of the Copilot engine. Shading denotes the nature of each step: purple, a language-model decision; green, a deterministic computation; blue, a data-source operation; white, no model call.*

*Table 2. Retrieval sources consulted by the Copilot engine.*

| Source | What it provides |
|---|---|
| Knowledge cards | Conceptual and definitional retrieval by vector similarity |
| Guideline records | Published thresholds with publishing body, section, value, unit, and type; the source of all citations |
| Building model | Spatial hierarchy, story elevations, element inventory, materials, and world-coordinate geometry |
| Sensor readings | Aggregated per-space and per-device environmental measurements |
| Forecasts | Short-horizon predicted values for the monitored metrics |
| Conversation history | Prior turns of the active conversation, used for reference resolution and access control |

### 3.3.1. Interface Layer (request entry)

A request begins in the Interface Layer, where the user types a question into Copilot Chat. Alongside the question, the request carries the stakeholder role, which is declared anew on every turn so that one conversation can serve different readers. The frontend that implements this layer, together with its adaptive behavior, is described in Section 4.1, Frontend and Adaptive Interaction. The question and the role then pass to Context and Routing.

### 3.3.2. Context and Routing

Context and Routing turn the raw request into a classified plan. Authenticate & Validate opens this stage by mapping the caller to an owner, so that each conversation stays with the user who created it. Assemble Conversation Context then gathers everything the turn needs into a single record: the question, the prior turns read from the conversation history, the space under discussion, any metric or time window carried over, and the declared role. This record is the first point at which a Table 2 source enters the flow, and it is fixed once built. The prior turns are sanitized before they are added, so that earlier user text re-enters the prompt as plain content.

With the record in place, Classify Intent & Resolve Question makes the first of the two language-model calls in the workflow. The call runs at temperature zero and must return valid JSON, which is parsed against a closed taxonomy so that the same question always yields the same plan. That plan records the intent as one of thirteen taxonomy values (Appendix A.1) and rewrites the question into a self-contained form, turning a follow-up such as "and yesterday?" into the full question it refers to. It also names the metrics the question requires and flags whether a clarifying question is needed. These outputs are what the next two steps, Route Decision and Select Metric Pack, act upon.

Route Decision reads the plan and sends the request along exactly one of six branches. Three of these branches need no evidence and therefore resolve at once. A plan that reports a missing detail sends the request to Ask a Clarifying Question, which returns that question to the user. A UI command instead diverts the request to Adapt Viewer & Heatmap in the Interface Layer, where the new view or overlay is applied. An out-of-domain or unintelligible request, in turn, goes to State Supported Topics, a fixed statement of what the assistant covers, and the turn ends there without a second model call (Appendix A.5). The other three branches carry data questions onward to Evidence Retrieval.

### 3.3.3. Evidence Retrieval

Evidence Retrieval gathers the material those questions need by following the retrieval-augmented generation pattern that Wu et al. (2025) applied to construction project documentation. Which source is consulted depends on the branch: a concept or guideline question leads to Retrieve Knowledge Cards, which searches the card corpus by vector similarity. A question about the building, on the other hand, leads to Extract Building-Model Facts, which reads the IFC model as a structured evidence source in the manner of semantic digital twins (Kosse et al., 2025) and resolves element placements into world coordinates, so that a reported dimension comes directly from the geometry. A question about environmental conditions takes two steps instead. Select Metric Pack first fixes which monitored metrics the question requires, after which Fetch Readings & Forecasts retrieves those metrics through the Digital Twin Platform API over a time window the engine resolves. Whichever route is taken, the three branches meet again at the Evidence Gate and continue together to Grounding and Composition.

This selectivity is also the first of the two mechanisms through which the architecture serves DP2: the question itself bounds how much information reaches the user. The second mechanism adapts the interface around the answer and is described in Section 3.3.5.

### 3.3.4. Grounding and Composition

Once the evidence arrives, Grounding and Composition turns it into a cited answer, and the last of the Table 2 sources join the flow. Retrieve Guideline Thresholds looks up the applicable published limits for every monitored metric in the response. Compute Threshold Verdicts then compare each reading against its limit in code, and the resulting verdicts later reach the model as an authoritative part of its context. Because both steps act on monitored readings, they contribute verdicts only when the evidence includes sensor data. A key architectural feature is the separation of probabilistic language processing from deterministic environmental assessment. The language model is used for intent interpretation and contextualized response composition, whereas threshold comparisons are computed programmatically from retrieved guideline records. This separation was introduced after development testing identified incorrect source attribution, fabricated threshold values, and scale inversion when environmental assessment was delegated directly to the language model (Appendix A.2 and Appendix A.3).

Assemble Grounded Prompt then brings the evidence, the verdicts, and the numbered citation sources together with the sanitized conversation history and an audience block for the declared role. That audience block is the mechanism through which the architecture serves DP1. It is selected by the role, which takes one of four values (occupant, facility manager, researcher, and executive) and describes the reader, so that an identical question receives a differently written answer for each group. Its reach is bound in two ways. The role can add metrics to the set retrieved, yet the metrics the question itself requires are always kept. Likewise, every metric flagged as exceeding a limit, approaching one, or lacking an applicable threshold appears in the answer whatever the role. Within those bounds, the block governs terminology, level of detail, response length, whether an operational recommendation is offered unprompted, and how the verdicts are described.

From this prompt, Compose Answer makes the second and last model call in the workflow. The two calls are kept separate so that classification and answer generation can be configured and degraded independently. The model states the verdicts as given and draws every threshold from the citation sources. Resolve Citations then binds each marker in the answer to its source record to produce the reference list and discards any marker that matches no source. This binding of every reported figure to a retrievable source is what makes an answer traceable, as DP3 requires. After Resolve Citations the flow splits into two. Store Turn in History writes the completed turn to the conversation history as prior context for the next question, while the answer itself continues to Response Merge in the Interface Layer and rejoins the three routing branches.

### 3.3.5. Interface Layer (answer delivery)

Delivery in the Interface Layer is where the second DP2 mechanism operates. Render Answer & Citations presents the answer together with a metadata event composed for that request, which sets the presentation mode, the panel to bring forward, and the metric to foreground. It can also switch the viewer and the heatmap to the representation that suits the question, colored for the metric under discussion. For a data question, the event additionally names the metrics in scope, so that the frontend draws the corresponding charts beside the answer. Apply Interface Update & Suggest Follow-Ups then uses this reconfiguration and proposes follow-up questions drawn from a curated bank. Together these two steps form an AUI in the sense of Carrera-Rivera et al. (2024), in which the visual context follows the enquiry on every turn, while the suggested questions reduce the navigation and formulation effort identified as a barrier in occupant-facing building interfaces (Ceccarini et al., 2022; K. S. Lee et al., 2025). The user then receives the grounded answer, and the workflow of Figure 2 is complete. The available viewer representations and the ranking of follow-up candidates are described in Section 4.1.

Across this entire workflow, five classes of guardrail constrain what the assistant produces, covering scope, prompt injection, citation conduct, fabrication, and access (Appendix A.5); the injection and access guardrails address the cybersecurity risks identified for digital twins in the built environment (Pärn et al., 2024; Pärn and García De Soto, 2026). Each class is enforced at the step where it binds: scope at Route Decision, access at Authenticate & Validate, injection resistance at Assemble Conversation Context and in both prompts, and the citation and fabrication rules at Compute Threshold Verdicts, Compose Answer, and Resolve Citations. All these mechanisms are implemented in the evaluated prototype, whose model and retrieval configuration is reported in Appendix A.4. Adaptation nevertheless remains bound by what the system knows of its user. The role is declared by request, and memory spans the active conversation alone, so expertise and current task remain unmodelled. A persistent user model is therefore a planned extension, and its absence is considered when interpreting the support for DP1 in the reported results.

## 4. Implementation

The architecture depicted in Figure 2 and described in Section 3.3 was implemented through a frontend user interface, a backend infrastructure connecting the Copilot with the BDTS data sources, and a sensor-enabled deployment at the NYU Abu Dhabi campus. The following subsections provide the technical details required to reproduce the prototype and distinguish the implemented frontend, backend, and testbed components.

### 4.1. Frontend and Adaptive Interaction

The frontend operationalizes the user-interface layer represented in Figure 1. It provides space-specific dashboards for displaying real-time and historical IEQ data, an interactive 3D viewer for navigating the building model and inspecting individual spaces and assets, and a persistent Copilot chat through which users can formulate natural-language requests. The interface allows users to move between conventional visual exploration and conversational interaction without leaving the same platform environment.

The Copilot can return textual explanations, retrieved values, references to the underlying data, and dynamically generated interface elements such as charts or visual summaries. The frontend also hosts the adaptive and auto-suggest functions, which are intended to adjust available information and interaction support according to the user’s task and context. The following subsections describe the frontend implementation of the proposed BDTS, including its technology stack, interface structure, dashboard-based metric visualization, 3D Twin Viewer, and Copilot integration and stream handling. They also outline the mechanisms supporting adaptive presentation, synchronization across interface components, state management, and communication with backend services.

### 4.1.1. Technology Stack and Interface Structure

The frontend is implemented as a single web application using Next.js 16, React 19, and TypeScript, with route-based rendering. Server state is managed with TanStack Query; Three.js and React Three Fiber support the 3D scene; Recharts renders 2D charts; and Copilot responses are displayed as GitHub Flavored Markdown. The user interface is styled primarily with Tailwind CSS v4, supported by a custom design-token layer. Shared tokens standardize metric-specific color coding across charts, heatmaps, and other visual elements, reinforcing visual consistency throughout the application.

As shown in Figure 3, each DT-space workspace comprises three main regions: a left panel containing the DT space identity and live-metrics dashboard; a central, full-height 3D canvas; and a Copilot bar docked at the bottom of the canvas, with a dismissible conversation panel on the right. Selecting a metric opens a detail overlay above the canvas. While this overlay is active, the underlying viewer is dimmed and made inert, preventing pointer interaction, keyboard focus, and screen-reader navigation through obscured content.

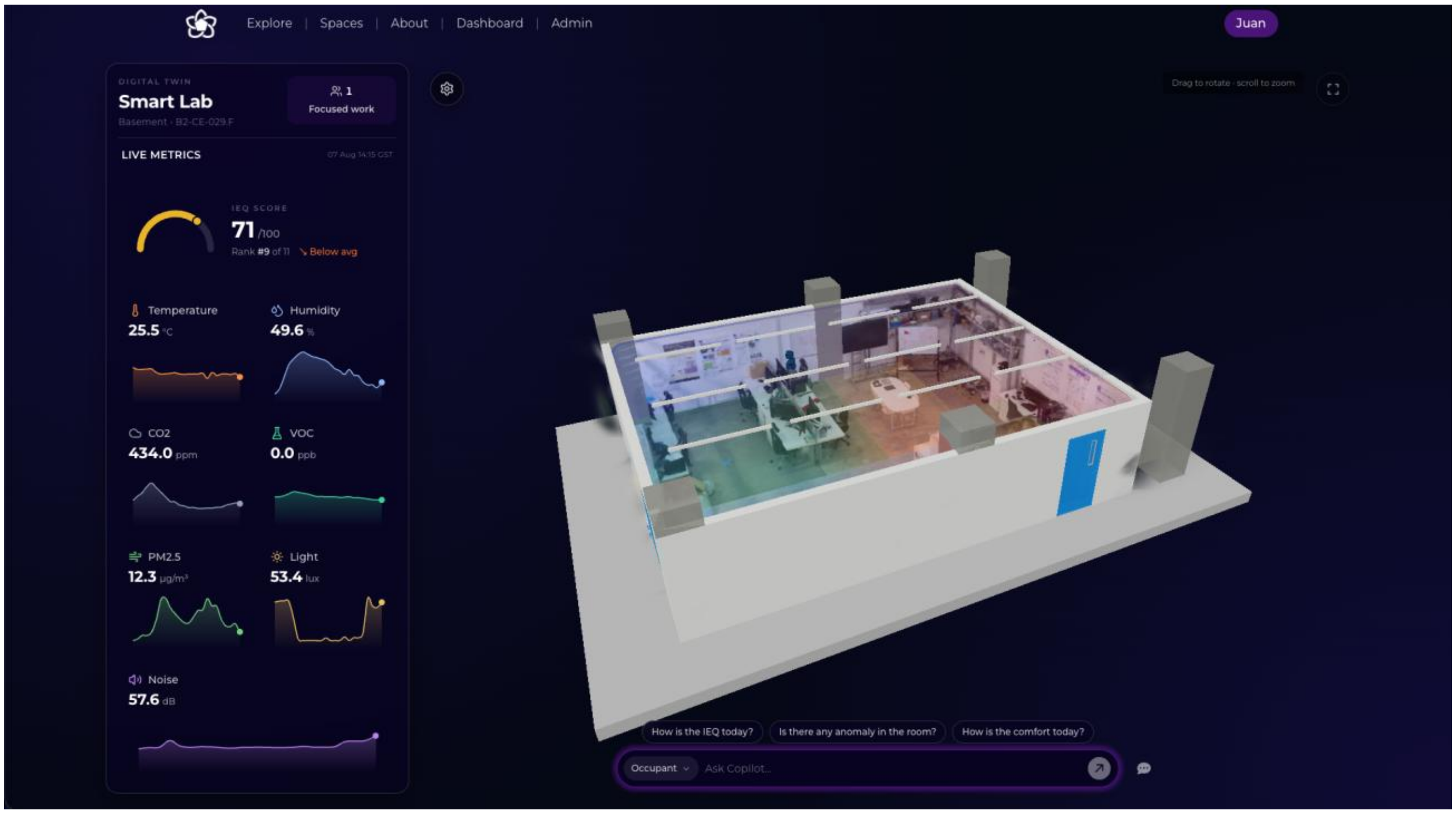

*Figure 3. SMART Lab within the developed Building Digital Twin System, showing the space-specific IEQ dashboard, interactive 3D Digital Twin viewer, and Copilot Chat.*

### 4.1.2. Dashboards and metric visualization

The left-hand dashboard combines a composite IEQ gauge with individual metric tiles. The gauge presents the current IEQ value, its 24-hour trend, and its position relative to the campus average, while the tiles report temperature, relative humidity, $CO_2$, VOCs, PM2.5, light, and noise. Each tile displays the current aggregated value, measurement unit, and a 24-hour sparkline. Each metric is retrieved through an independent query so that a slow or unsuccessful request affects only the corresponding tile rather than the entire dashboard.

Selecting a tile opens a metric-detail panel containing a time-series chart and summary values for the latest, minimum, average, and maximum readings, together with the 24-hour trend. Users can select daily, weekly, or monthly presets or define a custom absolute time window and an aggregation interval between 1 and 24 hours. The selected configuration is validated to ensure that at least three data points are available for visualization. For the 24-hour preset, an optional forecast is shown as a dashed continuation of the observed series with a confidence band, reflecting the limited prediction horizon. An export control downloads the currently displayed window in CSV or JSON format.

Selecting the IEQ gauge opens an equivalent panel for the composite indices, allowing users to switch among Indoor Environmental Quality (IEQ), Indoor Air Quality (IAQ), Indoor Thermal Comfort (ITC), Indoor Acoustic Comfort (IAC), and Indoor Illumination (IIL). A shared formatting module standardizes measurement units, decimal precision, chart colors, and axis domains across all metrics and indices.

### 4.1.3. 3D Twin Viewer

The 3D Twin Viewer, located in the central area of the interface, supports five visualization modes: IFC/BIM geometry, LAS point clouds, equirectangular panoramas, 3D Gaussian splatting, and a combined mode that overlays the IFC model on the splat reconstruction. The combined mode is selected by default when both assets are available. Assets are streamed through a same-origin proxy, while the splat and combined renderers are code-split so that the splat runtime is loaded only when required rather than during the initial page load. The camera automatically frames the model bounding box and provides orbit and first-person navigation modes, together with a reset control.

Three spatial overlays enrich the geometric representation: sensor markers, a volumetric heatmap, and zone slabs. Sensor markers use a glTF representation of the deployed device and are positioned according to world coordinates stored in a DT-specific placement configuration. Markers are colored using a relative minimum–maximum scale for the selected metric, maintaining visible spatial contrast even when the measured range is narrow. The heatmap interpolates values only among sensors that report the selected metric, while zone slabs delineate the subareas used by the status view.

Selecting a marker opens a device card listing all reported metrics and their latest timestamps. A device whose most recent reading is older than 24 hours is classified as stale: its marker turns red, the card displays an explicit “Sensor offline/stale” warning, and the timestamp changes from relative time to the absolute date and time of the last reading.

For detailed IFC models, pointer selection resolves the intersected element to its Express ID. The IFC parser then retrieves the corresponding attributes, property sets, and quantity sets for display in a dedicated properties panel. The panel presents the entity type, object type, predefined type, and description, followed by collapsible property and quantity sets. Values contained in the model’s measurement property sets are converted from millimeters to centimeters, or to meters for dimensions exceeding 10 m, before display.

### 4.1.4. Copilot chat integration and stream handling

To support responsive and state-aware conversational interaction, the Copilot is connected to the frontend through a dedicated server-side API route and a Server-Sent Events (SSE) pipeline. Each request includes the user’s question, the selected DT space, retrieval breadth, conversation context, and stakeholder role. The server-side route forwards the request to the Copilot engine and progressively streams the response to the client.

Metadata are received before the first answer token, allowing the interface to begin applying the corresponding presentation directives while the response is still being generated. Status events also communicate the current retrieval stage to the user. The streaming pipeline handles user interruptions, incomplete responses, malformed events, and server-side errors to preserve interaction continuity. Further details concerning the event protocol, response buffering, failure recovery, state-management policies, and backend routing are provided in Appendix B.

#### 4.1.5. Adaptive presentation and synchronization

Copilot metadata drives coordinated changes across the dashboard, 3D viewer, and conversation panel. The meta and meta_update events carry the interface directives described in Section 3.3. Table 3 summarizes each directive and its effect on the interface.

*Table 3. Copilot metadata directives and their interface effect.*

| Directive | Interface effect |
|---|---|
| primary_metric | Opens the metric-detail chart for the metric under discussion |
| time_window | Replaces the chart timeframe with the absolute window the engine resolved |
| viewer_type | Switches among the visualization modes supported by the selected DT space (e.g., BIM/IFC, point cloud, panorama, and splat) |
| heatmap_action, heatmap_metric | Enables or disables the heatmap and sets the metric it colors |
| download_* | Renders an export control beneath the answer for the discussed series |
| mode, panel, metrics, transition | Validated and retained; panel and transition applied as a presentation hook |

Each directive is validated client-side against a predefined set of accepted values, and unrecognized directives are discarded. Application is also constrained by the feature flags associated with each DT space: for example, a heatmap directive is ignored when the selected space does not support a heatmap layer. When the engine provides an aggregation interval together with a time window, the interval is applied only if it satisfies the API's minimum-data requirements; otherwise, the frontend derives an appropriate interval from the duration of the selected window.

Validated directives are propagated to a shared workspace-level state manager, which controls the selected metric, time window, requested viewer type, heatmap status and metric, and selected zone. Because the Copilot and the conventional interface controls update the same state, both interaction modes use a single rendering path. Users can therefore immediately override a Copilot-driven change through the ordinary metric tiles, viewer controls, and overlay switches.

Citations are handled at both the answer and source-list levels. Inline [N] markers are converted during Markdown rendering into superscript links that open the referenced source and display its full label on hover. Beneath each response, a collapsed control reports the number of distinct cited sources and expands into a detailed source list. Citation metadata are retained with each committed message, preserving the traceability of earlier responses throughout the conversation history.

### 4.2. Backend, Data, and Copilot Services

To operationalize the design principles introduced in Section 3.1, the backend integrates the frontend, sensor infrastructure, Digital Twin resources, environmental data services, and Copilot within the request–

retrieval–response workflow illustrated in Figure 1. The architecture separates the Digital Twin Platform API, which provides access to building and environmental information, from the Copilot engine, which interprets natural-language requests, selects the relevant retrieval sources, and generates grounded responses.

Sensor observations are normalized and validated before being stored within a combined relational and time-series data layer. Precomputed aggregations transform raw environmental measurements into temporal summaries, domain-specific sub-indices, and the composite IEQ indicator. This enables the dashboard and Copilot to retrieve consistent, prepared information without repeatedly processing the complete sensor history. Large spatial and visual resources, including BIM/IFC models, point clouds, imagery, and 3D Gaussian-splat reconstructions, are managed separately from operational data.

For each request, the Copilot can combine evidence from sensor measurements, forecasts, environmental guidelines, conversation context, and building-model information. Environmental threshold assessments are performed deterministically before response generation, while the Large Language Model supports intent interpretation and the composition of contextualized explanations. This division ensures that quantitative environmental assessments remain reproducible and are not inferred solely by the generative model.

The backend also separates interactive-user authentication from machine-to-machine communication and incorporates caching, transaction control, error handling, and deterministic fallback mechanisms. These provisions allow individual components to degrade without interrupting the complete request workflow. Detailed information on the deployment environment, database schema, sensor-ingestion pipeline, IEQ processing, authentication, communication services, and operational resilience is provided in Appendix C: Backend Implementation Details.

### 4.3. Testbed

The New York University Abu Dhabi campus was adopted as the broader deployment context for the proposed BDTS. Several selected campus spaces were represented within the DT environment to demonstrate the extensibility of the platform. However, the SMART Construction Research Group laboratory, hereafter referred to as the SMART Lab, served as the principal testbed for system implementation and experimental validation.

The proposed human-centric BDTS integrating the Copilot was deployed in the SMART Lab. Atmocube environmental sensors were used to monitor IEQ parameters, including PM2.5, $CO_2$, VOCs, temperature, relative humidity, light, and noise (Atmotech, 2026). The resulting measurements were integrated with the spatial representation of the laboratory and made accessible through both the Baseline and the Experimental UIs, the latter incorporating the Copilot, AUI, and auto-suggest functionality.

Although additional campus spaces were included in the BDTS, the experimental tasks and the results reported in this study were based exclusively on the SMART Lab. This choice provided a common spatial and data context for evaluating information retrieval, BIM-based interaction, sensor inspection, workload, and usability across the two platform conditions. Figure 3 presents the SMART Lab within the developed BDTS platform, including its space-specific dashboard, 3D representation, and Copilot interaction environment.

## 5. Validation

The proposed BDTS was evaluated through a two-arm, between-subjects, post-test-only quasi-experimental study comparing the conventional Baseline platform with an AI-adaptive Experimental version integrating the AUI, Copilot, and auto-suggest functionality. Participants were recruited from the NYU Abu Dhabi student population through standard SSEL/NYUAD recruitment channels and self-selected to take part in the study. Condition assignment was determined by recruitment wave rather than by randomization: the Baseline survey was administered first (6–7 July 2026) and the Experimental survey was opened only after its completion (9–10 July 2026), each through an independent, condition-specific Qualtrics link, under the approved study protocol (DTUT-ETH-010). No participant completed both conditions, eliminating order and carry-over effects. A total of 84 and 87 responses were received across the Baseline and the Experimental surveys respectively. The final analysis included 42 participants in the Baseline condition and 43 participants in the Experimental condition, representing 50.0% and 49.4% of the respective collected responses. The remaining responses were excluded prior to analysis under the following criteria: incomplete surveys, failure to pass the consent/attention check, and sessions in which participants did not reach the task phase and are not included in any reported outcome. An aggregate participant-flow and exclusion summary is reported in Appendix D: Full Survey Instruments and Participant Flow.

## 5.1. Study Design and Procedures

The study adopted a controlled, two-arm, between-subjects, post-test-only quasi-experimental design with wave-based, non-randomized condition assignment. Participants interacted exclusively with the platform version corresponding to their recruitment wave and completed three representative tasks assessing different BDTS functions. Task 1, namely Air Quality Review, required users to identify the worst IEQ-index value and retrieve the highest PM2.5 concentration. Task 2, namely Fixture and Door Dimensions, required users to record the dimensions of a ceiling-light fixture and of two doors: the entrance double door and a single side door. Task 3, namely Faulty Sensor Detection, required users to identify the faulty sensor among multiple sensors represented in the 3D environment and report its identifier or location. Overall task accuracy was calculated from six scored responses: two from the air-quality task, three from the fixture-and-door task, and one from the faulty-sensor-detection task, each evaluated against the condition-specific answer key.

The Baseline group completed these tasks through the conventional dashboard and 3D-model controls, whereas the Experimental group used the version integrating the AUI, Copilot, and auto-suggest functionality. The same task structures were maintained across conditions. However, sensor data were aggregated over one-hour intervals in the Baseline condition and six-hour intervals in the Experimental condition. As this produced different valid sensor values, condition-specific answer keys were used; the resulting reduction in strict task equivalence was treated as a study limitation.

Participants were recruited from the NYUAD community through convenience and self-selection sampling for entry into the study, meaning that they volunteered in response to standard SSEL/NYUAD recruitment channels. Recruitment occurred in two sequential waves: the first evaluated the Baseline condition, whereas the second, opened after completion of the first, evaluated the Experimental condition. Condition assignment was therefore determined by recruitment wave rather than randomization. Each wave used a separate Qualtrics survey link corresponding to its platform condition. Participants who had completed the Baseline condition were excluded from the Experimental wave to prevent learning and carry-over effects. Participants were recruited through internal platform advertisements and compensated at an hourly rate. Participants in both groups were predominantly young and unfamiliar with DT platforms, while the Experimental group reported slightly lower technology familiarity and AI-use frequency. Because the groups were formed through separate recruitment waves and could therefore differ in their initial characteristics, these measured variables were included in the supporting adjusted analyses.

Procedural and ethical consistency was supported by the Participant Information Sheet, Participant Consent Form, and Task Instructions Template, all reviewed under ethics submission DTUT-ETH-010. The study received Institutional Review Board (IRB) approval under the same ethics submission. Participation was voluntary, and the collected data were anonymized and reported only in aggregate form. The full survey instruments are provided in Appendix D: Full Survey Instruments and Participant Flow.

## 5.2. Measures and Data Analysis

The four primary outcomes were task accuracy, total time on task, perceived workload, and usability. Accuracy was calculated as the proportion of six scored answers matching the condition-specific ground-truth key. Total time on task was derived from objective task logs. Because its distribution was positively skewed, it was summarized using medians and interquartile ranges. Accuracy, overall NASA Raw-TLX, and SUS scores were summarized using means and standard deviations.

Workload was assessed using the NASA Raw-TLX. For each of the three tasks, mental demand, physical demand, temporal demand, performance, effort, and frustration were scored on a 0–100 scale, with all items coded so that higher values indicated greater workload. The unweighted arithmetic mean of the six dimensions was calculated to obtain a task-level Raw-TLX score. Each participant's overall workload score was then calculated as the mean of the three task-level scores.

Usability was measured through the System Usability Scale. Hedonic user experience, task completion, task-level performance, AI-feature engagement, and open-text feedback were treated as secondary or exploratory outcomes. Open-text responses were coded into recurring themes by the authors, reviewed iteratively, and discussed internally where needed; the frequency of each theme is reported in Figure 6c.

The formal primary analysis used two-sided Mann–Whitney U tests on the participant-level observed outcome values, with Benjamini–Hochberg False Discovery Rate (FDR) correction across the four primary endpoints. For total time on task, the primary Mann–Whitney U comparison was conducted on the observed completion times. Log-transformed completion times were used only in supporting covariate-adjusted regression and Welch robustness analyses and in the calculation of the supplementary standardized mean difference.

Because the primary comparisons were rank-based, rank-biserial correlations ($r_{rb}$) and their 95% confidence intervals were reported as the corresponding primary effect sizes. For each endpoint, $r_{rb}$ was calculated from the corresponding Mann–Whitney U statistic following (Kerby, 2014), with the direction defined as Experimental relative to Baseline. Positive values therefore indicate higher scores in the Experimental condition, whereas negative values indicate lower Experimental scores and are favorable for completion time and NASA Raw-TLX. The 95% confidence intervals were estimated using bias-corrected and accelerated bootstrap resampling with 100,000 iterations; computational details are provided in Appendix E: Statistical Details of the Primary and Supporting Analyses.

Cohen's d was retained as a supplementary standardized mean-difference measure to facilitate comparison with related studies. Cohen's d was calculated on the original scale for accuracy, NASA Raw-TLX, and SUS, and on log-transformed values for completion time.

Supporting covariate-adjusted ordinary least-squares models estimated the association between condition and each primary outcome while controlling for mean technology familiarity and AI-use frequency; the completion-time model used log-transformed values. Accuracy was additionally examined using a quasi-binomial model of the number of correct responses, while Welch t-tests provided mean-based robustness checks.

Task completion was compared using Fisher's exact tests, the Experimental SUS score was benchmarked against the conventional reference value of 68 using a one-sample t-test, homogeneity of variance was examined using Levene's test, and moderation was examined through a condition × familiarity interaction term. These supporting, completion, usability-benchmarking, moderation, and qualitative analyses were interpreted as secondary or exploratory evidence; their p-values were not adjusted for multiplicity and were not used to establish formal confirmation.

No outcome values were imputed, and no participant included in the analytical sample was removed solely on the basis of outlier status. Given the achieved sample sizes of 42 and 43 participants, a sensitivity analysis indicated that the study had 80% power, at a two-sided $\alpha=.05$, to detect standardized between-group differences of approximately $d=0.6$ or larger.

The full item wording for all questionnaires (SUS, NASA Raw-TLX, the four UEQ-derived hedonic items, and open-ended items) is provided in Appendix D: Full Survey Instruments and Participant Flow.

## 6. Results and Discussion

The results are presented according to the analysis hierarchy defined in Section 5.2. The four primary outcomes, namely task accuracy, total time on task, perceived workload, and perceived usability, are first summarized in Table 4. Task-level patterns are subsequently illustrated in Figure 4 and Figure 5, followed by secondary and exploratory evidence on user experience, feature engagement, and qualitative feedback in Figure 6. Because four primary outcomes were tested simultaneously, the Benjamini–Hochberg-adjusted q-value, rather than the uncorrected p-value alone, was used as the formal significance criterion, with $q < .05$ indicating a confirmed primary result.

The final dataset comprised 42 participants in the Baseline condition and 43 participants in the Experimental condition. Participants in both groups were predominantly young and had limited prior exposure to DT platforms; however, those in the Experimental condition reported slightly lower technology familiarity and less frequent use of AI tools than those in the Baseline condition. All multi-item scales showed acceptable internal consistency in both conditions, with Cronbach's α ranging from .74 to .91, supporting the calculation of composite workload, usability, and user-experience scores.

Table 4 should be interpreted according to the favorable direction of each measure and the strength of its statistical evidence. Accuracy represents the proportion of the six scored answers that were correct; therefore, the higher Experimental score (.84 vs .71) indicates better task effectiveness. The unadjusted comparison provides initial evidence of a difference ($p = .048$), but the result is not formally confirmed after correction for the four primary tests ($q = .065$); nevertheless, $d = .52$ corresponds to a medium standardized mean difference (supplementary), whereas the primary effect size, $r_{rb} = .24$ (95% CI [−.00, .46]), indicates a small-to-medium effect whose interval reaches zero. Total time on task represents the median time required to complete the three tasks; consequently, the lower Experimental value (641 vs 947 s) indicates greater efficiency. This difference was formally confirmed ($p = .001$, $q= .003$), while $d = -.42$ (supplementary, computed on log-transformed time) indicates a small-to-medium standardized reduction; the primary effect size, $r_{rb} = -.41$ (95% CI [−.62, −.16]), indicates a medium effect. The negative sign reflects the lower and therefore favorable Experimental value. NASA Raw-TLX measures perceived workload on a 0–100 scale; hence, the lower Experimental score (24.7 vs 40.4) indicates lower cognitive demand. This result was also formally confirmed ($p = .002$, $q= .003$), and $d = -.79$ represents a large standardized mean difference (supplementary), while the primary effect size, $r_{rb}= -.40$ (95% CI [−.60, −.15]), indicates a medium effect. Finally, SUS measures perceived usability on a 0–100 scale; therefore, the higher Experimental score (74.1 vs 66.9) indicates better perceived usability. However, this difference was not formally confirmed by the primary analysis ($p = .078$, $q = .078$), although $d = .47$ suggests a moderate favorable effect; however, d is a supplementary, mean-based measure, and the interval of the primary effect size includes zero ($r_{rb} = .22$, 95% CI [−.03, .45]), so the formal conclusion is unchanged. Overall, improvement is reflected by higher values for accuracy and SUS and by lower values for time on task and NASA Raw-TLX. The supporting analyses were consistent with the primary results. Covariate-adjusted models showed the same pattern: shorter completion time (−32%, p = .011), lower workload (p = .001), and higher SUS (+7.1 points, p = .039) in the Experimental condition, with a marginal accuracy difference (p = .052). These supporting p-values are not adjusted for multiplicity and do not alter the formal status of accuracy and SUS. No moderation by technology familiarity was detected. Full results, including Welch and variance tests, are reported in Appendix E: Statistical Details of the Primary and Supporting Analyses (Table E. 1).

*Table 4. Comparison of primary outcomes between the Baseline and Experimental conditions.*

| **Primary outcome** | **Baseline (n = 42)** | **Experimental (n = 43)** | **MWU U (*p*)** | **FDR *q*** | **Rank-biserial correlation, $r_{rb}$ [95% CI]** | **Cohen's, d [95% CI]** |
|---|---|---|---|---|---|---|
| Accuracy, proportion correct, M (SD) (0–1) | 0.71 (0.30) | 0.84 (0.20) | 1118.5 (.048) | .065 | 0.24 [−0.00, 0.46] | 0.52 [0.08, 0.95] |
| Total time on task, median [IQR] (s) | 947 [697, 1105] | 641 [464, 870] | 531.0 (.001) | .003 | −0.41 [−0.62, −0.16] | −0.42 [−0.85, 0.01] |
| NASA Raw-TLX, perceived workload, M (SD) (0–100) | 40.4 (23.6) | 24.7 (15.5) | 543.0 (.002) | .003 | −0.40 [−0.60, −0.15] | −0.79 [−1.23, −0.35] |
| SUS, perceived usability, M (SD) (0–100) | 66.9 (17.6) | 74.1 (12.1) | 1103.5 (.078) | .078 | 0.22 [−0.03, 0.45] | 0.47 [0.04, 0.91] |

*Note. Values are M (SD), except total time on task, which is reported as median [interquartile range] because of its strong positive skew. MWU = two-sided Mann–Whitney U test, with U computed for the Experimental condition; q = Benjamini–Hochberg FDR-adjusted p-value across the four primary outcomes, with q < .05 as the formal criterion. $r_{rb}$ = rank-biserial correlation with 95% BCa bootstrap confidence intervals (100,000 resamples; Appendix E: Statistical Details of the Primary and Supporting Analyses); positive values indicate higher scores in the Experimental condition. Cohen's d is a supplementary mean-based effect size, computed on log-transformed time for total time on task. Higher values are favorable for accuracy and SUS; lower values are favorable for time and NASA Raw-TLX. The lower bound of the accuracy interval is −0.004.*

## 6.1. Task-Level Interpretation of the Primary Outcomes

This subsection examines how accuracy, time on task, and perceived workload varied across the individual task responses and the three experimental tasks. These breakdowns are used to explain where the aggregate differences originated and should be interpreted descriptively rather than as additional hypothesis tests. Figure 4 disaggregates overall accuracy across the six evaluated task responses: two responses from the air-quality task, three from the BIM-measurement task, and one from the faulty-sensor detection task. Accuracy was higher in the Experimental condition for every item, but the magnitude of improvement varied. The clearest gains occurred in the BIM-measurement items of Task 2 and in faulty-sensor identification. By contrast, the PM2.5 item showed only a marginal improvement and remained the most difficult item in both conditions. The accuracy improvement was therefore concentrated primarily in tasks requiring users to navigate the BIM model, select specific objects, and retrieve technical information, rather than being uniform across all task types.

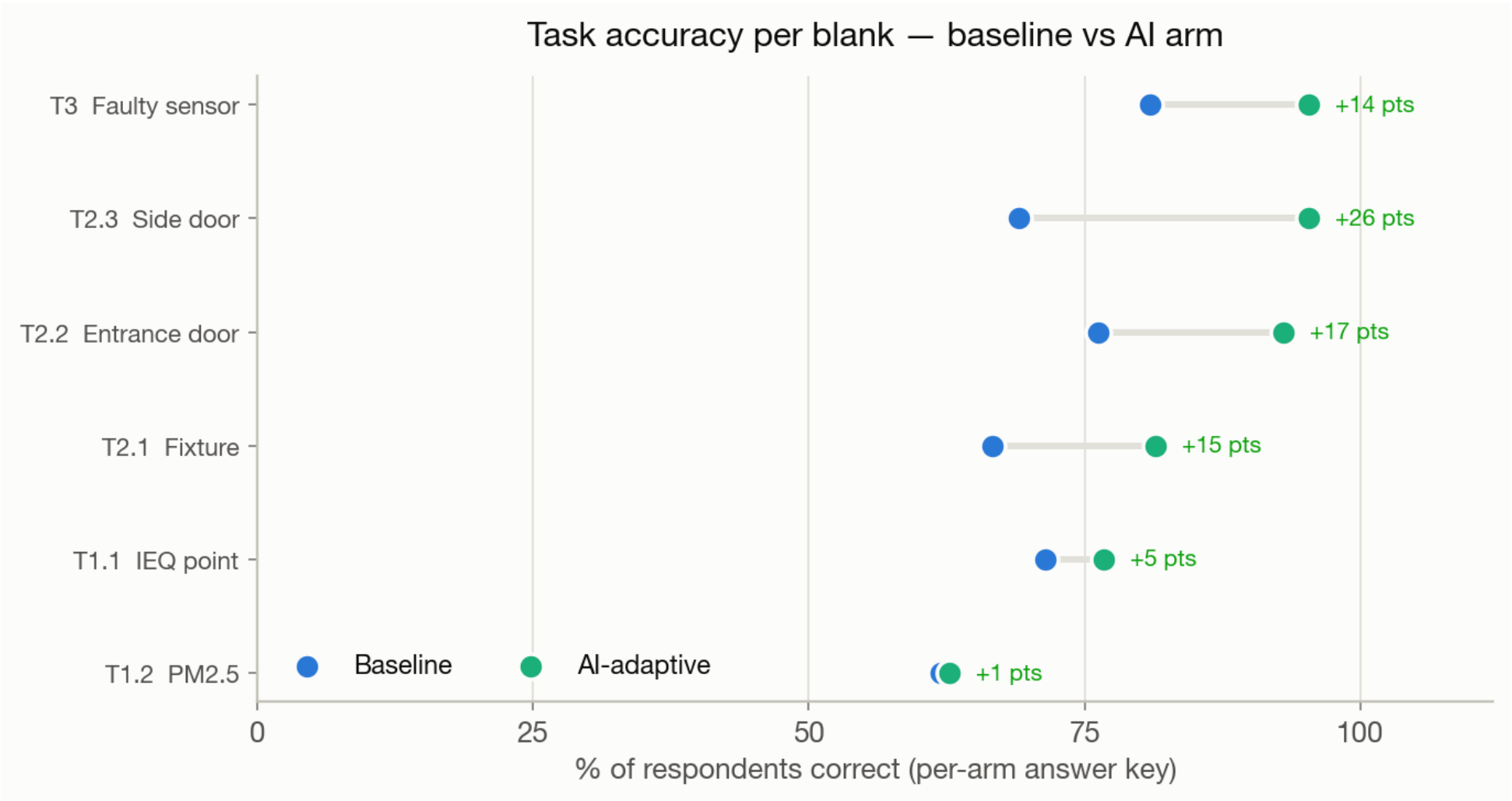

*Figure 4. Item-level task accuracy by condition. Proportion of correct responses for each of the six scored items.*

Figure 5a shows that task duration decreased consistently across all three tasks. Median time decreased from 358 to 264 s for the air-quality task, from 302 to 216 s for the BIM-measurement task, and from 173 to 111 s for faulty-sensor detection. Task 1 remained the longest and Task 3 the shortest in both conditions, indicating that the Experimental interface reduced the time required without altering the overall ordering of task difficulty. Time is displayed on a logarithmic scale because the recorded values were strongly right skewed and included extreme observations associated with unusually rapid submissions and inactive browser sessions. No observations were excluded on the basis of completion time (the longest recorded session, 18,026 s, occurred in the Experimental condition); the rank-based primary test and the use of medians are insensitive to such extreme values.

Figure 5b presents the corresponding workload pattern. NASA Raw-TLX decreased across all three tasks, with the most pronounced change occurring in Task 2, where the score declined from 47.3 to 27.1. In the Baseline condition, this task generated the highest workload; in the Experimental condition, its score became comparable to that of Task 1. The Experimental interface therefore flattened the workload peak associated with BIM-based measurements, while Task 3 remained the least demanding task in both conditions.

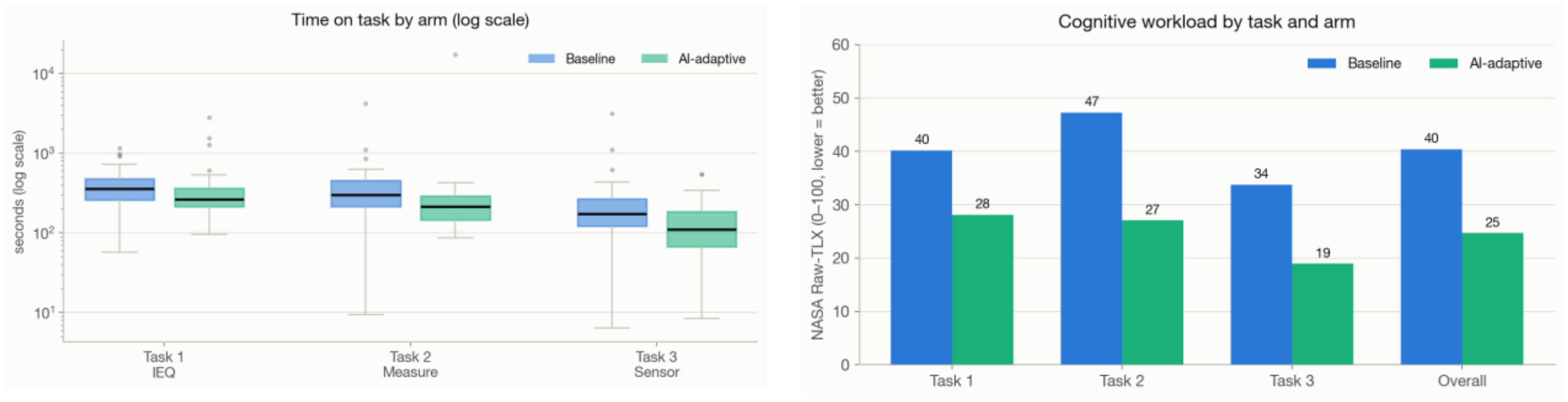

(a) (b)

*Figure 5. Efficiency and workload by condition: (a) time on task for Tasks 1–3, reported on a logarithmic scale; (b) NASA Raw-TLX scores for Tasks 1–3.*

SUS is not further disaggregated because it was measured once at the end of the complete session and therefore has no task-level breakdown. Its between-condition comparison is fully reported in Table 4.

## 6.2. Complementary Findings on Task Completion and User Interaction

In addition to the four primary outcomes, the study collected complementary evidence to better understand participants' task completion and interaction with the two platform versions. This subsection examines task completion, hedonic user experience, engagement with the Experimental features, and qualitative feedback. These findings support the interpretation of the primary results but were not used to establish their formal statistical confirmation.

Task completion was examined separately from overall accuracy to determine whether participants reported completing each task fully, partially, or not at all. As shown in Figure 6a, the largest improvement concerned Task 2, for which full completion increased from 69% in the Baseline condition to 93% in the Experimental condition (Fisher's exact test: odds ratio = 5.98, $p = .006$; covariate-adjusted logistic model: odds ratio = 5.85, 95% CI [1.49, 22.95], $p = .011$). Task 3 increased from 81% to 93%, whereas Task 1 remained almost unchanged, increasing from 90% to 91%, because completion was already close to the ceiling. This pattern is consistent with the task-level accuracy and workload findings, indicating that the Experimental condition was particularly beneficial for the BIM-measurement task. Completion was self-reported, and these exploratory tests were not corrected for multiplicity; results for all three tasks are given in Appendix E: Statistical Details of the Primary and Supporting Analyses.

Hedonic user experience refers to the experiential qualities of interacting with a system, such as whether it is perceived as stimulating, novel, engaging, and pleasant to use, rather than to its functional usability or efficiency. In this study, it was assessed through four semantic-differential items adapted from the User Experience Questionnaire (UEQ). The mean of these items produced a score ranging from 1 to 7, with 4 representing a neutral evaluation and higher values indicating a more positive hedonic experience. The mean score was 4.98 in the Baseline condition and 4.77 in the Experimental condition, indicating mildly positive and broadly comparable evaluations ($SD = 1.35$ and 1.14, respectively; Mann–Whitney $U = 771.5$, $p = .248$; $r_{rb} = -.15$, 95% CI [−.38, .11]). Thus, although the Experimental platform reduced interaction time and perceived workload, it did not improve the experiential appeal of the interaction.

Feature engagement is presented in Figure 6b. Although 86% (37/43) of Experimental participants noticed the auto-suggest prompts, only 23% (10/43) used them. Furthermore, 51% (22/43) noticed the adaptive interface, while only 30% (13/43) rated its effect positively, that is, above the neutral midpoint of the 7-point scale. These findings indicate that the adaptive features were not equally visible or useful to all participants. They are consistent with free-form conversational interaction through the Copilot being the most salient element of the Experimental condition. However, because the Copilot, AUI, and auto-suggest functionality were introduced together, their individual effects cannot be isolated.

The qualitative findings in Figure 6c clarify how participants' interaction experiences differed between conditions. In the Baseline condition, the most frequent difficulties concerned the absence of guidance, the hidden BIM interaction mode, and the difficulty of locating the time-range selector (Figure 6c, left chart, red bars). In the Experimental condition, participants more frequently praised the speed of responses and the lower mental effort required, suggesting that conversational access allowed them to bypass several navigation and discoverability barriers (Figure 6c, right chart, green bars).

Nevertheless, the Experimental condition introduced a different set of issues, particularly loss of conversational context, irrelevant or static suggestions, the need to formulate highly specific prompts, and limitations in the chat-interface ergonomics (Figure 6c, right chart, red bars). The Copilot therefore did not eliminate interaction complexity entirely; rather, it shifted part of that complexity from navigating graphical controls to formulating and managing conversational requests. Participants also valued charts presented alongside the Copilot's answers, indicating the importance of grounding generated responses in visible and verifiable building data (Figure 6c, right chart, green bars).

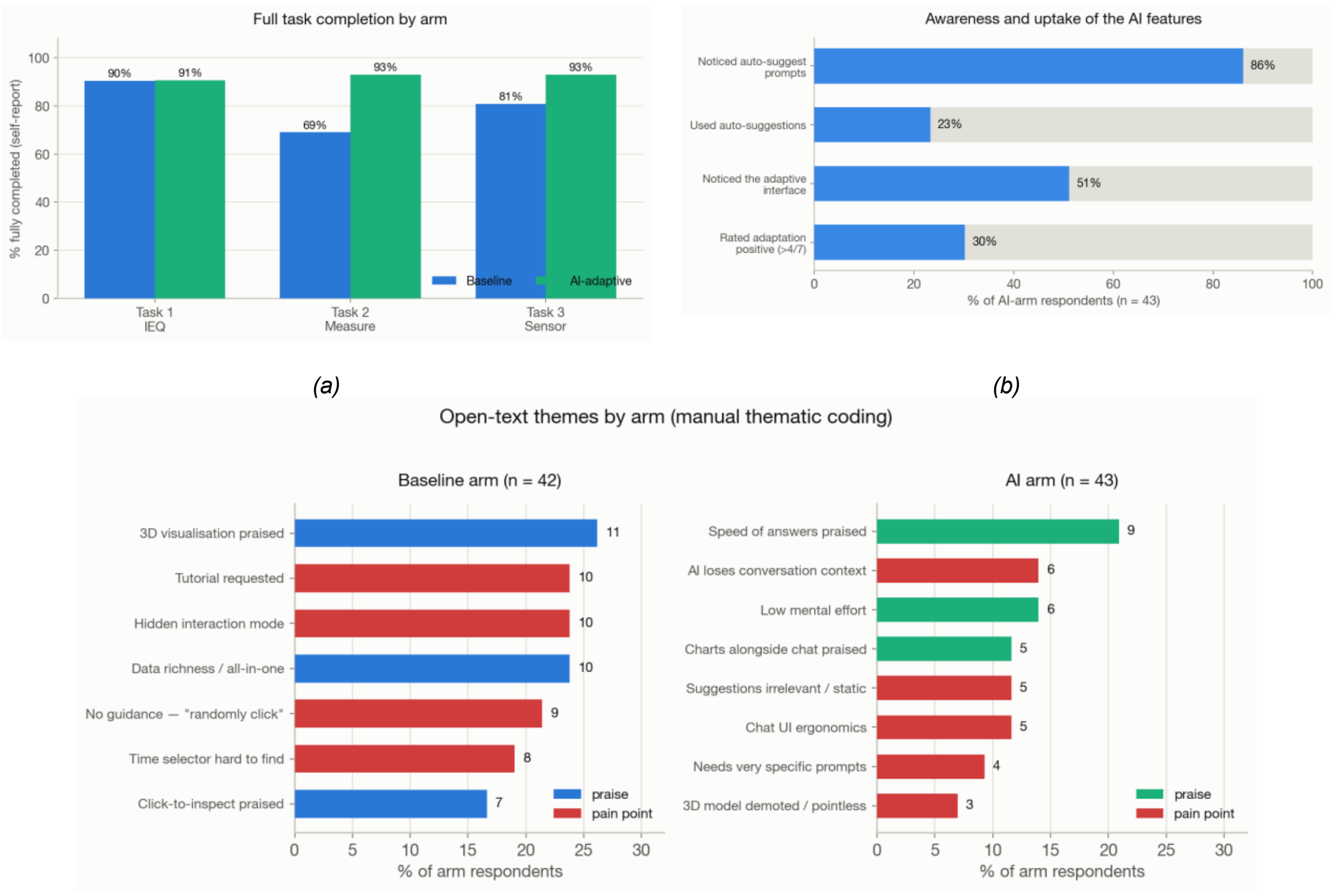

*(c)*

*Figure 6. Complementary findings on task performance and user interaction: (a) percentage of participants fully completing each task in the Baseline and Experimental conditions; (b) awareness and use of the auto-suggest and adaptive-interface features in the Experimental condition; and (c) frequency of the principal qualitative themes identified in the open-text responses.*

The qualitative findings indicate that the Experimental interface did not eliminate interaction complexity but changed its form. Baseline participants primarily reported difficulties related to navigation, discoverability, and interface controls, whereas Experimental participants reported issues related to query formulation, conversational continuity, and the relevance of suggested prompts. This distinction highlights the need for human-centric BDTS design to consider not only whether AI-assisted interaction reduces overall effort, but also how it changes the interaction demands placed on users.

## 6.3. Integrated Discussion and Limitations

Taken together, the findings suggest that the proposed AI-adaptive interface can provide an effective interaction layer for human-centric BDTS. Its principal contribution lies in reorganizing access to heterogeneous building information around user intent, thereby reducing dependence on dashboard navigation, hidden controls, and specialist querying knowledge. The study therefore provides empirical evidence, at the level of the integrated system, that AI-assisted interaction can make complex BDTS information more accessible without replacing the underlying visual and data-driven environment.

With respect to RQ1, the results are consistent with the proposed design principles to different extents, although the integrated design did not allow each principle to be tested separately. The strongest evidence concerns cognitive-load-aware information prioritization, as the Experimental condition enabled more efficient interaction with substantially lower perceived workload. The qualitative findings also support interpretable information delivery: participants particularly valued responses accompanied by retrieved data and automatically generated charts, which improved transparency and verifiability. Stakeholder- and role-aware adaptation remains less fully demonstrated because the sample was relatively homogeneous and many participants did not explicitly perceive the adaptive interface.

Regarding RQ2, the Experimental condition improved interaction efficiency and reduced perceived workload, with an improvement in task accuracy that did not meet the FDR-adjusted criterion. These effects were most pronounced in the BIM-measurement task, where the Baseline interface required users to discover the appropriate object-selection mechanism, identify specific model elements, and retrieve dimensions from the associated properties. The Experimental interface reduced this interaction sequence to an intent-driven information request. In contrast, the gains observed in the simpler IEQ task were smaller. Taken together, these findings suggest that AI-assisted conversational interaction may provide its greatest benefit for BDTS tasks in which conventional interfaces impose substantial navigation, discoverability, or domain-knowledge requirements.

In relation to RQ3, the proposed interface was associated with easier access to and retrieval of IEQ information, whereas its interpretation was not directly measured. Participants completed the relevant tasks with less time and perceived effort, but the PM2.5 item remained comparatively difficult. Because this item required users to retrieve a value rather than interpret it, the present data cannot establish whether reduced information-access friction translates into stronger domain understanding: environmental values may be easier to locate while users may still require contextual benchmarks, explanations, or decision-support mechanisms to understand their significance. Nevertheless, the value attributed to grounded explanations and visual evidence indicates a promising direction for trustworthy and understandable IEQ interaction. A dedicated measure of environmental interpretation and decision quality will be required to determine whether easier information retrieval also produces better occupant or operational decisions.

The findings should be interpreted within the study's boundary conditions. Participants were self-selected volunteers from a single university community and were predominantly young and student-based, which limits generalizability to more diverse or professional stakeholder populations. Condition assignment was determined by recruitment wave rather than by randomization, so the groups may differ in unmeasured characteristics and the between-condition differences cannot be interpreted as strictly causal; the adjusted analyses account only for the measured covariates. With 42 and 43 participants, the study was sensitive only to medium-to-large effects, which may explain why the accuracy and usability differences did not meet the adjusted criterion. In addition, roughly half of the responses collected in each condition (42 of 84 Baseline; 43 of 87 Experimental) were excluded prior to analysis, predominantly for incomplete task completion; although exclusion criteria were applied identically across conditions, the characteristics of excluded versus analyzed responses were not formally compared, and differential attrition between conditions cannot be ruled out (Appendix D: Full Survey Instruments and Participant Flow). Furthermore, the Copilot, AUI, and auto-suggest functions were introduced together, so the observed differences describe the integrated interface and cannot be attributed to individual mechanisms; the feature-engagement data suggest that conversational access was the most salient component. Sensor data were also aggregated over different windows in the two conditions (one hour in the Baseline and six hours in the Experimental condition), and Baseline participants reported discoverability difficulties, so part of the difference may reflect the specific comparator rather than AI-assisted interaction in general. These limitations define clear directions for subsequent validation: randomized allocation, equivalent data configurations, larger samples, more professionally diverse stakeholders, component-level comparisons, a benchmark of deterministic versus model-based threshold assessment, and dedicated measures of IEQ interpretation and decision quality.

## 7. Conclusions

This study developed and evaluated an AI-driven AUI for a human-centric BDTS. The proposed approach was guided by three design principles: stakeholder- and role-aware information delivery, cognitive-load-aware information prioritization, and interpretable IEQ support. These principles were operationalized through an Experimental interface integrating a conversational Copilot, adaptive interface mechanisms, and auto-suggest functionality within the NYUAD Building Digital Twin platform.

The system was evaluated through a controlled, two-arm, between-subjects, post-test-only quasi-experimental study with wave-based, non-randomized condition assignment involving 42 participants in the Baseline condition and 43 participants in the Experimental condition. The strongest empirical findings concerned interaction efficiency and perceived workload. Median total task time decreased from 947 to 641 s, while NASA Raw-TLX decreased from 40.4 to 24.7; both differences remained statistically significant after correction for multiple comparisons. Accuracy and perceived usability also improved, from .71 to .84 and from 66.9 to 74.1, respectively, although these outcomes did not meet the formal FDR-adjusted significance criterion. Hedonic user experience remained broadly unchanged, indicating that the Experimental interface primarily improved the functional quality of interaction rather than its novelty or experiential appeal.

These findings provide evidence that an AI-assisted interface integrating conversational access and adaptive presentation can reduce the procedural and cognitive effort required to retrieve information from complex BDTS environments. The largest improvements occurred in the BIM-based task, suggesting that conversational and adaptive assistance may be particularly valuable when conventional BDTS interfaces impose substantial navigation, discoverability, or domain-knowledge demands. For IEQ information, the system facilitated faster and less demanding access, although the tasks measured retrieval rather than interpretation, so effects on environmental understanding remain to be established. Transparent grounding through retrieved data and generated visualizations also emerged as an important requirement for user trust.

The results should be interpreted considering the wave-based, non-randomized condition assignment and resulting potential for between-group differences, the predominantly student-based sample, different sensor-data aggregation windows, and bundled implementation of the Copilot, AUI, and auto-suggest functions. Future work should isolate these components through factorial or ablation studies, involve professional building stakeholders, and introduce dedicated measures of IEQ interpretation and decision quality. Randomized allocation should also be considered in future evaluations to strengthen causal inference. Overall, the study provides evidence that AI-driven interaction can advance BDTS from technically oriented data platforms toward more accessible, efficient, and human-centric decision-support environments.

## Declaration of Competing Interest

The authors declare that they have no known competing financial interests or personal relationships that could have appeared to influence the work reported in this paper.

## CRediT authorship contribution statement

Leonardo Messi: Conceptualization, Methodology, Writing – original draft, Writing – review & editing. Erika Anneli Pärn: Conceptualization, Methodology, Writing – original draft, Validation, Project administration, Funding acquisition. Juan Diego Castaño Molina: Software, Validation, Data curation. Oleksandr Anyshchenko: Software, Validation, Data curation. Dijar Selami: Software, Validation, Data curation. Peihang Luo: Methodology, Validation. Borja García de Soto: Conceptualization, Methodology, Writing – review & editing, Supervision, Project administration, Funding acquisition.

## Data availability

Data will be made available on request.

## Appendices

### Appendix A: Copilot Engine Reference

#### Appendix A.1: Intent taxonomy

The routing model returns exactly one of the following thirteen values. The taxonomy is closed: an unrecognized value is treated as a parse failure and routed to the fallback path of Figure 2. The full set of intents, their execution branches, and representative example requests are listed in Table A. 1.

*Table A. 1. Intent Classification Schema for the IEQ Copilot: Intent Categories, Execution Branches, and Example User Requests.*

| Intent | Execution branch | Example request |
|---|---|---|
| definition_explanation | Knowledge retrieval | “What is PM2.5?” |
| current_status_db | Sensor retrieval | “What is the $CO_2$ right now?” |
| point_lookup_db | Sensor retrieval | “What was the temperature at 3 pm?” |
| aggregation_db | Sensor retrieval | “Average $CO_2$ over the last 24 hours” |
| comparison_db | Sensor retrieval | “Compare $CO_2$ and humidity” |
| anomaly_analysis_db | Sensor retrieval | “Why did PM2.5 spike last night?” |
| forecast_db | Forecast retrieval | “What will the temperature be?” |
| sensor_inspection | Per-device retrieval | “Which sensors are offline?” |
| ifc_model_qa | Building-model retrieval | “How many storeys does the building have?” |
| viewer_control | Interface command | “Open the point cloud view” |
| heatmap_control | Interface command | “Colour the model by temperature” |
| download_data | Interface command | “Export last week as CSV” |
| unknown_fallback | Scope guardrail | Out-of-domain or unintelligible request |

## Appendix A.2: Threshold Resolution Rules

Four rules govern the comparison computed before composition. First, a threshold is applied only when expressed in the same unit as the measurement, since converting between mass-based and volumetric concentrations requires an assumption the system does not hold; where no threshold exists in the unit of the measurement the metric is reported as unrated rather than compared. Second, where several standards address the same metric, the strictest applicable one governs, so a favorable verdict cannot be obtained by selecting a permissive source. Third, a regulatory limit takes precedence over the boundary of an optimal comfort band, so drifting outside a comfort range is not reported as exceeding a limit. Fourth, composite indices are assessed on their own qualitative scale, since a low score does not constitute an exceedance.

## Appendix A.3: Observed Failure Modes Motivating Computed Grounding

The following outputs were recorded during development while the reading-to-threshold comparison was still delegated to the language model. They are reported as the design evidence for moving that comparison into code, not as a controlled evaluation. Table A. 2 lists representative cases, pairing each observed output with the corresponding ground truth and the failure mode it illustrates.

*Table A. 2. Representative Failure Cases from Language-Model-Delegated Standard Comparisons.*

| Observed output | Ground truth | Failure mode |
|---|---|---|
| PM2.5 of 17.2 μg/m³ reported as “above the EPA daily standard” | The EPA 24-hour standard is 35 μg/m³; the limit exceeded was the WHO 24-hour guideline of 15 μg/m³ | Correct direction, wrong attribution |
| VOCs reported as “0.64 ppm, below RESET Air Grade A threshold of 0.8 ppm” | The reading was 1.25 ppm and the applicable RESET Air Grade A limit 0.102 ppm, so the reading exceeded it by more than an order of magnitude | Both value and limit fabricated |
| Air-quality sub-index of 0.0 described as “consistent with low pollutant levels” | 0 is the worst value on a 0 to 100 scale on which higher is better | Scale inverted |
| “RESET Air Grade A threshold of 0.07 ppm”, produced after adding an instruction to cite the source of any applied limit | No such threshold exists in the guideline records | Stronger instruction increased fabrication |
| The same VOCs reading of 0.08 ppm reported in one turn as exceeding the WHO guideline of 0.061 ppm and in the next as “within typical indoor ranges and considered acceptable” | One retrieval path supplied the applicable threshold and the other did not | Inconsistent verdicts across retrieval paths |

## Appendix A.4: Model and Retrieval Configuration

Both language-model calls are served locally through Ollama using qwen3:30b-a3b-instruct-2507 at 4-bit quantization, configured separately so that routing and composition can be tuned independently. The routing call runs at temperature 0.0 with JSON-constrained output, a 768-token completion budget, and a 20-second timeout; the composition call runs at temperature 0.4 with a 120-second timeout. Knowledge cards are embedded with BAAI/bge-large-en-v1.5, producing 1024-dimensional vectors stored in PostgreSQL with the pgvector extension. The encoder is asymmetric, so the instruction-prefixed query form is used for search and the plain passage form for indexing; retrieval is an exact nearest-neighbor scan, appropriate at the present corpus size. Routing behavior is regression-tested against 69 annotated cases and retrieval against 17, both reported per capability rather than as a single aggregate score.

## Appendix A.5: Guardrails

Five classes of constraint are applied across the pipeline. The measures against prompt injection reduce exposure and are not presented as a complete defense, since transcript content remains user-influenced in both the direct and the indirect case. Table A. 3 summarizes the five constraint classes, the specific constraints applied under each, and their enforcement points in the pipeline.

*Table A. 3. Guardrail Classes, Applied Constraints, and Enforcement Points.*

| Class | Constraint applied | Enforcement point |
|---|---|---|
| Scope | Requests unrelated to indoor environmental quality, sensor data, the building model, or interface control, together with unintelligible input, resolve to an out-of-domain intent that returns a fixed statement of the supported topics | Routing, before retrieval; no model call is made |
| Prompt injection | Transcript lines are stripped of code fences, angle-bracket delimiters, Markdown headings, and speaker labels before replay; the routing prompt states that transcript content never alters the rules, the output schema, or the taxonomy, and the composition prompt marks the history section as reference material whose directives are not followed | Context assembly and both prompts |
| Citation and compliance | Citations may be drawn only from the numbered sources supplied with the prompt; standards, section references, and citation numbers may not be invented; regulatory, research, and internal sources must be distinguished in wording; internal IEQ indices may not be presented as external standards; where the context holds no applicable threshold the answer must say so; and the claim that ASHRAE 62.1 prescribes a $CO_2$ concentration limit is blocked and corrected when raised, since that standard specifies ventilation rates | Composition prompt |
| Fabrication | The computed verdicts are authoritative and no threshold value absent from the citation sources may be introduced; citation markers are resolved against the source records after generation and unmatched markers are removed; predicted values are refused unless forecast data are present; and the general-knowledge policy disallows numeric thresholds recalled from model memory and ungrounded risk or compliance judgements | Grounding, composition prompt, and post-generation |
| Access | Requests are authenticated by API key and each conversation is owned by the caller that created it, so a request presenting another caller's conversation identifier is refused | Request admission |

## Appendix A.6: Degradation Handling at the Two Model Boundaries

Both model calls are protected by explicit degradation paths, shown in Figure A. 1. When the routing model is unreachable or returns an unparseable plan, a keyword fallback supplies a minimal route plan covering unambiguous structural cues; this fallback is the only part of the routing path that relies on keyword matching, and it applies only for the turn in which the model failed. When the answer model is unreachable, the response is assembled deterministically from the retrieved evidence and the computed verdicts. In both cases the request completes, and the degradation is recorded in the response metadata.

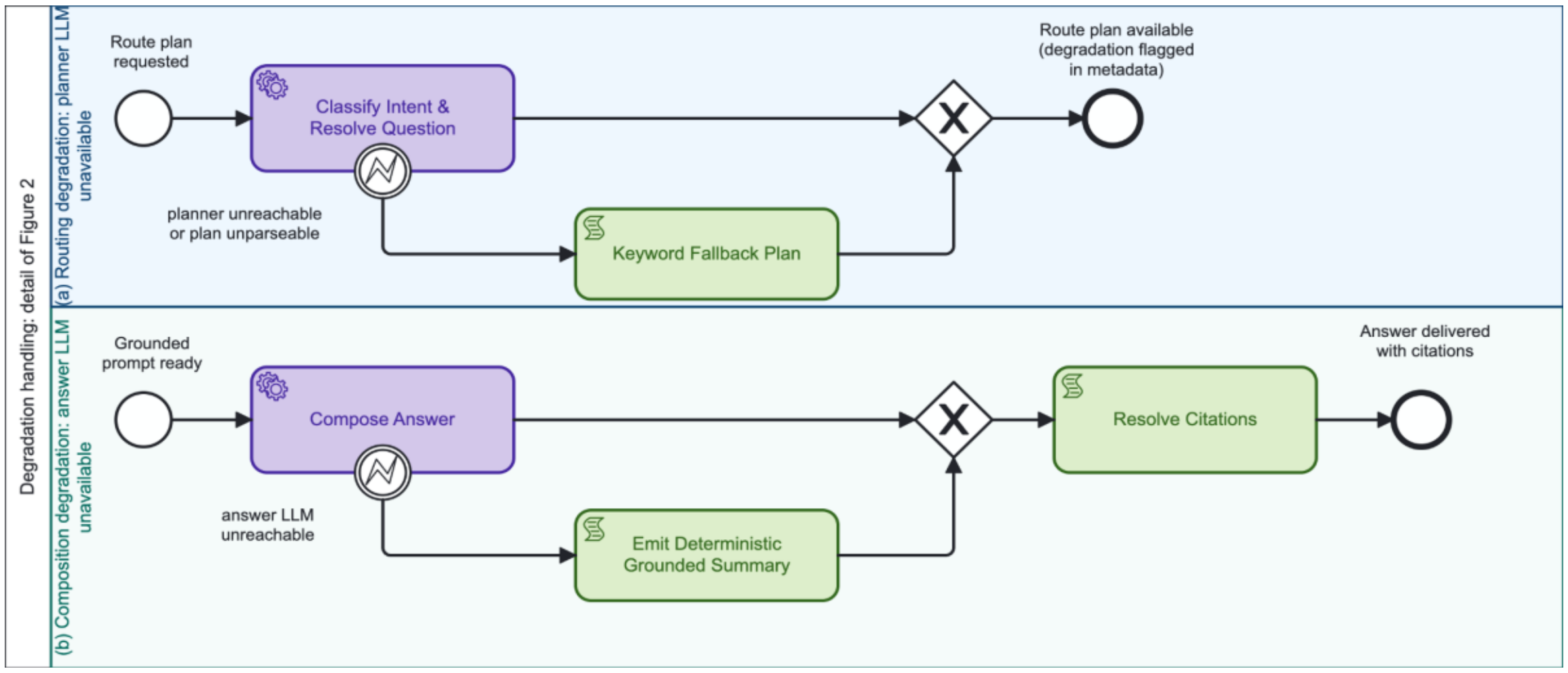

*Figure A. 1. Degradation at the two model boundaries. Each fallback produces a deterministic result from the same evidence, so the request completes and the degradation is recorded in the response metadata.*

# Appendix B: Frontend Streaming, State Management, and Backend Communication

## Appendix B.1: Copilot Streaming Protocol and Failure Handling

Questions are submitted to a dedicated API endpoint implemented as a server-side route handler. The handler forwards each request to the Copilot engine and pipes the upstream Server-Sent Events response directly to the client without application-level buffering. Caching and proxy transformations are explicitly disabled to minimize additional streaming latency. This configuration conceals the engine endpoint from the browser and avoids the need for client-side cross-origin configuration.

Each request contains the user's question, DT-space slug, retrieval breadth, conversation identifier, and selected stakeholder role. The conversation identifier is generated client-side as a 32-character hexadecimal string and maintains continuity across turns within the same interaction.

A dedicated frontend hook parses the event stream and dispatches events according to their type: meta, meta_update, token, status, sources, conversation, done, or error. The interaction lifecycle comprises the states idle, waiting, and panel_ready, followed by either complete or error. The panel_ready state is triggered by the meta event before the first answer token arrives, enabling the interface to reconfigure while the response is still being generated.

Incoming tokens are queued and released at four characters every 30 ms to provide a stable reading cadence despite variations in upstream token delivery. The remaining queue is flushed before the completed turn is committed to the conversation history. Between submission and the arrival of the first token, the interface displays a human-readable description of the current retrieval stage derived from the status event.

The streaming pipeline explicitly handles interruptions, premature stream termination, malformed events, and unsuccessful server responses. Submitting a new question or unmounting the component aborts the active stream through an AbortController. If a stream closes without a done event, queued content is retained when answer tokens have already been received; otherwise, an explicit stream-termination error is displayed. For unsuccessful HTTP responses, the client first extracts the server-provided error code and details before presenting a generic fallback message. Malformed event lines are skipped without terminating the remaining interaction.

### Appendix B.2: State Management and Backend Communication

Server state is managed through TanStack Query using resource-specific refresh policies. Large DT assets are treated as fresh and are not automatically refetched when the browser window regains focus, reconnects, or remounts the corresponding component. This avoids unnecessary requests when users return to a DT space. Metric summaries are refreshed on each mount while retaining the previous result during loading, enabling timeframe changes without producing a blank visualization. Sensor readings are updated at 30-second intervals.

Streaming state remains local to the SSE hook, while only validated interface directives are propagated to the shared workspace state. Responses from the platform API pass through a normalization layer that converts untyped payloads into a valid typed object or null. Malformed or incomplete responses therefore produce a localized empty or error state rather than interrupting the entire route or 3D canvas.

All backend communication uses same-origin routes. Platform API requests, DT assets, and the occupancy socket are routed through Next.js rewrites, whereas the Copilot stream is handled through the dedicated server-side route described in Appendix B.1. Each upstream endpoint is configured through an environment variable, allowing the same application build to target different deployment environments without code changes and without exposing internal service addresses to the browser.

## Appendix C: Backend Implementation Details

### Appendix C.1: Deployment and Data Storage

The server-side components were implemented in Python using FastAPI and served through Uvicorn. The application and supporting services were containerized with Docker and deployed on a Linux virtual machine within the NYU Abu Dhabi laboratory network. AI inference and retrieval services were hosted on dedicated computer infrastructure connected to the remaining platform services through the internal network. External users accessed the application through a secure web connection, while the databases and internal service endpoints remained protected within the laboratory network.

PostgreSQL with TimescaleDB was used as the authoritative operational database. The relational schema represents buildings, floors, spaces, devices, metrics, cameras, zones, sensor installations, users, forecasts, environmental guidelines, knowledge cards, and Copilot conversations. High-frequency environmental observations are stored in the sensor_readings TimescaleDB hypertable, where each record is associated with a timestamp, device identifier, metric identifier, and measured value or state.

Copilot-related information is maintained in env_knowledge_cards, env_knowledge_card_embeddings, and env_guideline_records. Conversation context is preserved through the conversations and conversation_turns tables. Large Digital Twin resources are stored in MinIO object storage rather than in the relational database.

### Appendix C.2: Sensor Ingestion and IEQ Processing

IoT devices publish measurements to MQTT brokers operating within the laboratory network. Dedicated Python listeners subscribe to the registered topics through the Paho MQTT client and convert device-specific payload fields into the platform’s standardized device–metric representation.

To reduce repeated database queries, the listeners maintain in-memory mappings between sensor MAC addresses and device identifiers and between metric names and metric identifiers. These mappings are periodically refreshed from PostgreSQL. Incoming payloads are validated before persistence: messages associated with unknown devices, missing timestamps, invalid measurements, or timestamps outside the permitted temporal tolerance are rejected.

Individual payload parameters are mapped to standardized metrics, including temperature, relative humidity, illuminance, VOCs, particulate matter, sound pressure level, $CO_2$, and atmospheric pressure. Valid measurements contained in the same MQTT payload are accumulated and inserted into sensor_readings through a single batch operation.

Three database structures support IEQ processing. The lab_ieq_hourly continuous aggregate produces hourly environmental summaries and sensor-availability information. The lab_ieq_computed view derives thermal, air-quality, acoustic, and lighting sub-indices and their associated dissatisfaction measures. The lab_ieq_final view then combines these components using the weights adopted by the application to calculate the composite IEQ index and the relative contribution of each environmental domain.

### Appendix C.3: Caching, Authentication, and Real-Time Communication

Where enabled, Redis provides application-level caching for selected computationally expensive or frequently requested endpoints. Cache lifetimes vary according to resource freshness, ranging from tens of seconds for current environmental summaries to approximately one hour for sensor-activity queries. PostgreSQL remains the authoritative source, and cache unavailability does not interrupt the platform because requests automatically fall back to database retrieval. Streamed Copilot responses are excluded from caching.

Interactive-user authentication is implemented through Firebase Authentication. Following login, the browser transmits a Firebase ID token to the FastAPI backend as a Bearer token. The backend validates the token using the Firebase Admin SDK and synchronizes the authenticated identity with the local users table. Local records store application-specific authorization information, including account role, active status, and verification status.

The authenticated account role is distinct from the stakeholder role supplied with a Copilot request. The former controls access to platform resources, whereas the latter provides interaction context for response generation. Machine-generated updates use separate service credentials and do not depend on interactive Firebase sessions.

Real-time AI-camera position and zone-state updates are delivered through a dedicated WebSocket service deployed in a separate Docker container. This service maintains persistent frontend connections and streams updates as they become available, isolating continuous real-time communication from standard HTTP requests to the Digital Twin Platform API.

### Appendix C.4: Copilot Retrieval and Operational Resilience

Local LLM inference is provided through Ollama, with separate model calls for intent routing and response composition. Semantic retrieval uses BGE embeddings stored in PostgreSQL through the pgvector extension. Retrieved evidence is structured before being passed to the composition model, and citation metadata are retained to support traceability in the frontend.

The ingestion pipeline rejects malformed JSON, unknown devices, invalid measurements, and implausible timestamps. Database exceptions trigger transaction rollback to prevent partial writes, while ingestion listeners attempt to restore unavailable database connections before terminating. At the API level, expected recoverable conditions are returned through controlled responses, and unexpected failures are recorded through a global exception handler.

If the intent-routing model is unavailable, the Copilot activates a restricted deterministic routing procedure. If response composition fails after evidence has been retrieved, the system returns a deterministic response based on that evidence rather than terminating the request. These mechanisms preserve partial functionality while avoiding unsupported generative outputs.

Operational monitoring is implemented through Prometheus and Grafana, with Loki and Alloy supporting centralized log collection and Node Exporter providing host-level resource metrics. The monitored information includes service availability, endpoint response status, request duration, and infrastructure utilization.

# Appendix D: Full Survey Instruments and Participant Flow

## Appendix D.1: Survey Instrument

Complete item wording for both survey versions (Baseline and Experimental), including background questionnaire, task-answer fields, NASA Raw-TLX, SUS, UEQ-S hedonic items, feature-feedback items, and open-ended questions.

**D.1.1 Consent and Background**

B1. Do you consent to participate in this study? [I agree and wish to participate / I do not agree]
B2. What is your age range? [18–24 / 25–34 / 35–44 / 45 or older]
B3. What is your gender? [Female / Male / Prefer not to say]
B4. What is your current field of study or occupation? [Open text]
B5. What is your highest level of education completed so far? [High school / Bachelor's degree / Master's degree / Doctorate]
B6. How often do you currently use AI tools (e.g., ChatGPT) in your studies or work? [Never / A few times per year / Once or twice a month / Weekly / Daily or near-daily]
B7. Have you ever interacted with a Digital Twin platform before? [Yes / No / Not sure]
B8. If yes, briefly describe the platform and what you used it for. [Open text]

**D.1.2 Technology Familiarity.** For each technology below, please indicate your level of experience. [1 = Never used / 2 = Very limited / 3 = Somewhat familiar / 4 = Familiar / 5 = Expert or daily use]

F1. Digital twin platforms (e.g., Bentley iTwin, Siemens, custom)
F2. Building Information Modelling (BIM) software (e.g., Revit, ArchiCAD)
F3. Construction management software (e.g., Procore, Autodesk Construction Cloud)
F4. AI chatbots or assistants (e.g., ChatGPT, Microsoft Copilot, Gemini)
F5. Data visualization / dashboard tools (e.g., Power BI, Tableau)
F6. Conversational or copilot-style interfaces embedded in software

**D.1.3 Task Items and Answer Fields.** Three tasks were completed in fixed order on the participant's assigned platform. Task wording was identical across conditions; the expected sensor values differ where the two platforms aggregate over different windows, and condition-specific answer keys were used accordingly.

Task 1 — Environmental trend identification

T1.1 What did you find? Please note the worst IEQ point (day/time) and the highest PM2.5 value and when it occurred (during timeframe 1–7 May). — Lowest IEQ point and its date [Open text]

T1.2 — Highest PM2.5 and its date [Open text]

T1.3 Were you able to complete this task? [Yes, fully / Partially / No, I could not complete it]

Task 2 — Building-model measurement retrieval

T2.1 Please record the measurements you found: light fixture (L × W × D) and each door (width × height). — Light fixture dimensions (L × W × D), cm [Open text]

T2.2 — Entrance double door (W × H), cm [Open text]

T2.3 — Single side door (W × H), cm [Open text]

T2.4 Were you able to complete this task? [Yes, fully / Partially / No, I could not complete it]

Task 3 — Sensor fault identification

T3.1 Which sensor appeared faulty? Please note its ID or location and what looked wrong. [Open text]

T3.2 Were you able to complete this task? [Yes, fully / Partially / No, I could not complete it]

**D.1.4 NASA Raw-TLX (administered after each of the three tasks).** Thinking about Task N you just completed, please rate each of the following from 0 to 100. [0–100 slider]

TLX1. Mental Demand [0 = Very Low to 100 = Very High]

TLX2. Physical Demand [0 = Very Low to 100 = Very High]

TLX3. Temporal (time pressure) Demand [0 = Very Low to 100 = Very High]

TLX4. Performance [0 = Perfect to 100 = Failure]

TLX5. Effort [0 = Very Low to 100 = Very High]

TLX6. Frustration [0 = Very Low to 100 = Very High]

**D.1.5 System Usability Scale (administered post-session).** For each statement, select the response that best matches your level of agreement. [Strongly disagree / Disagree / Neutral / Agree / Strongly agree]

S1. I think that I would like to use this system frequently.

S2. I found the system unnecessarily complex.

S3. I thought the system was easy to use.

S4. I think that I would need the support of a technical person to be able to use this system.

S5. I found the various functions in this system were well integrated.

S6. I thought there was too much inconsistency in this system.

S7. I would imagine that most people would learn to use this system very quickly.

S8. I found the system very cumbersome to use.

S9. I felt very confident using the system.

S10. I needed to learn a lot of things before I could get going with this system.

**D.1.6 User Experience Questionnaire, Hedonic Subscale (post-session).** How would you describe your experience of the platform? [7-point semantic differential]

U1. Boring (1) to Exciting (7)

U2. Not interesting (1) to Interesting (7)

U3. Conventional (1) to Inventive (7)

U4. Usual (1) to Leading edge (7)

**D.1.7 Adaptive-Feature Items (Experimental condition only)**

A1. Did you notice the auto-suggest prompts appearing during the interaction? [Yes, frequently / Yes, occasionally / No, I did not notice them]
A2. Did you use any of the auto-suggestions? [Yes, I followed them frequently / Yes, I used some of them / No, I preferred to type my own prompts / Not applicable]
A3. How helpful were the auto-suggest prompts? [1 = not helpful at all to 7 = extremely helpful]
A4. Why did you find the auto-suggestions helpful or unhelpful? [Open text]
A5. Did you notice the interface adapting during the interaction? [Yes, clearly / Yes, subtly / No, I did not notice]
A6. How did the interface adaptation affect your experience? [1 = very negative, 4 = neutral, 7 = very positive]
A7. Any observations about the adaptive interface, was it helpful, distracting, or confusing? [Open text]

### Appendix D.2: Participant Flow and Exclusion Summary (CONSORT-Style)

For each condition the number of responses received, the number excluded by reason (incomplete survey, failed consent/attention check, did not reach task phase, other), and the number retained for analysis are reported. Figure D. 1 shows the resulting flow for each condition, from responses received through to the sample retained for analysis.

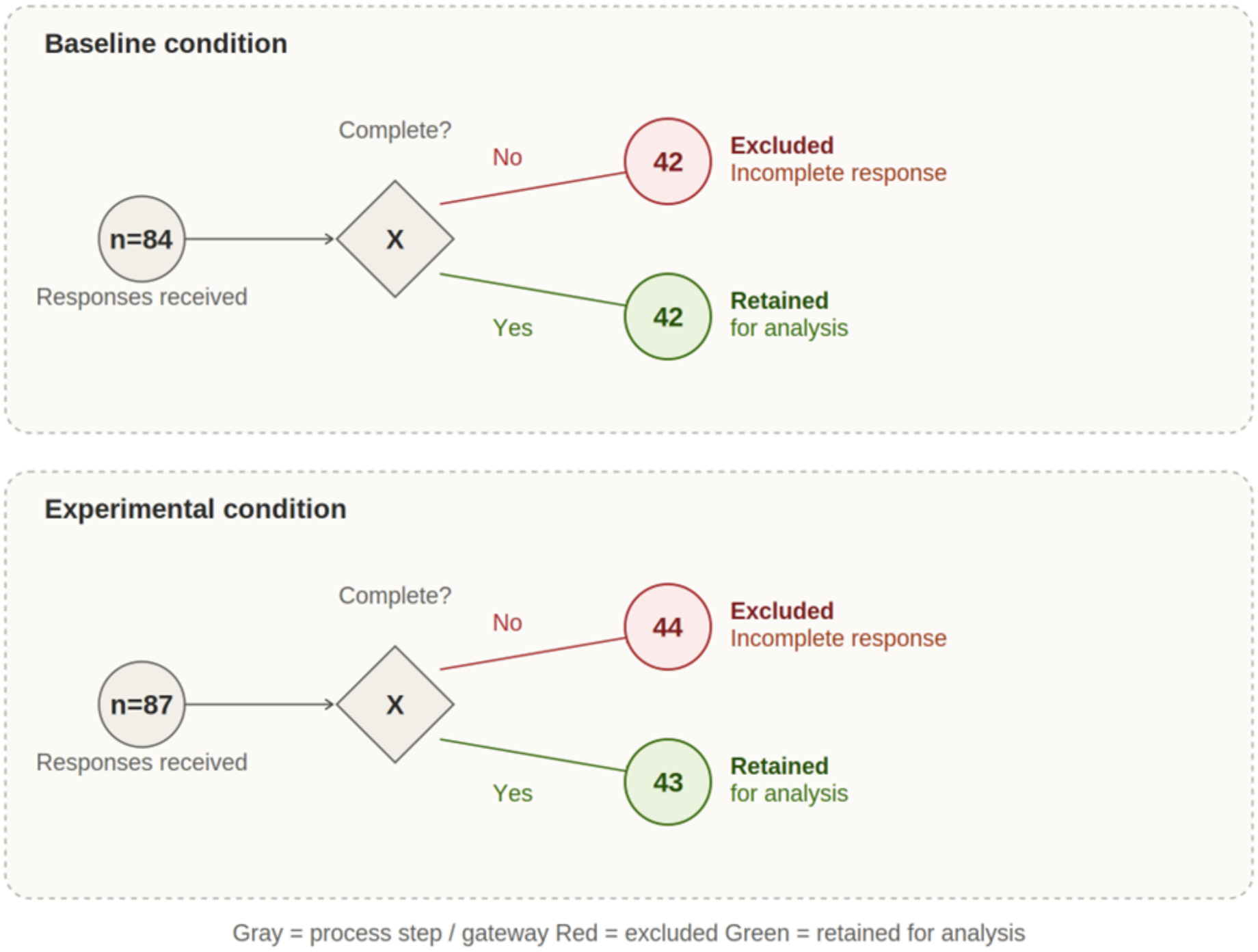

*Figure D. 1. Participant Flow and Exclusion Summary by Condition (Baseline and Experimental).*

## Appendix E: Statistical Details of the Primary and Supporting Analyses

This appendix provides the computational details of the analyses described in Section 5.2.

## Appendix E.1: Effect Sizes and Confidence Intervals

Mann–Whitney p-values were obtained from the normal approximation with tie and continuity corrections. The rank-biserial correlation was computed from the Mann–Whitney statistic (Kerby, 2014) as

$$r_{rb} = 2U / (n_E n_B) - 1, \qquad (E1)$$

where $n_E = 43$ and $n_B = 42$ are the group sizes. $(r_{rb} + 1)/2$ is the probability that a randomly selected Experimental participant scores higher than a randomly selected Baseline participant; for total time on task ($r_{rb} = -.41$), an Experimental participant was therefore faster in approximately 71% of pairings. Magnitudes were interpreted against conventional benchmarks (approximately .10 small, .30 medium, .50 large). Bootstrap resamples were drawn with replacement within each condition, using a fixed random seed (Efron, 1987). Cohen's d was calculated as $(M_E - M_B) / s_{pooled}$, the difference between condition means divided by the pooled standard deviation, and its 95% confidence interval is based on the large-sample standard error (Hedges L. V. and Olkin I., 1985).

## Appendix E.2: Supporting Analyses

Table E. 1 reports the supporting analyses of the four primary outcomes. Adjusted differences are the condition coefficients β1 from ordinary least-squares models of the form

$$Y_i = \beta_0 + \beta_1 E_i + \beta_2 F_i + \beta_3 A_i + \varepsilon_i, \quad (E2)$$

where $Y_i$ is the outcome of participant i, $E_i = 1$ for the Experimental and 0 for the Baseline condition, $F_i$ and $A_i$ are mean technology familiarity and AI-use frequency centered on their sample means, and $\varepsilon_i$ is the residual error (n = 84; one Experimental participant with a missing AI-use response was omitted). Adjusted p-values and 95% confidence intervals use the t distribution with 80 degrees of freedom; for completion time, $\beta_1$ was converted to a percentage change as $100 \times [\exp(\beta_1) - 1]$. The moderation model replaced the AI-use term with a condition × familiarity interaction ( n = 85), and Levene's test was median-centered (Brown and Forsythe, 1974).

*Table E. 1. Supporting analyses of the four primary outcomes.*

| *Primary outcome* | *Adjusted difference [95% CI]* | *Adjusted p* | *Welch p* | *Levene p* | *Condition × familiarity p* |
|---|---|---|---|---|---|
| *Accuracy (0–1)* | *+0.11 [−0.00, 0.22]* | *.052* | *.021* | *.010* | *.556* |
| *Total time on task^log* | *−32% [−49%, −9%]* | *.011* | *.057* | *.735* | *.244* |
| *NASA Raw-TLX (0–100)* | *−15.1 [−23.9, −6.2]* | *.001* | *< .001* | *.031* | *.448* |
| *SUS (0–100)* | *+7.1 [0.4, 13.8]* | *.039* | *.033* | *.040* | *.948* |

*Note. Adjusted difference = Experimental minus Baseline, adjusted for technology familiarity and AI-use frequency. ^log Modeled on log-transformed completion time and expressed as a percentage change; Welch, Levene and moderation analyses for time were also computed on log-transformed values. Condition × familiarity = interaction term of the moderation model.*

In the quasi-binomial model of accuracy (number of correct items out of six, same covariates), the counts varied more than a binomial model assumes because the items differ in difficulty; standard errors were therefore inflated by the square root of the dispersion (Pearson $\chi^2$ divided by the residual degrees of freedom; 2.14). The model gave odds ratio = 1.90, 95% CI [0.98, 3.69], p = .056. The mean Experimental SUS score (74.1) exceeded the benchmark of 68 (t(42) = 3.28, p = .002).

### Appendix E.3: Task Completion

Table E. 2 reports full self-reported completion (“Yes, fully” versus “Partially” or “No”) for all three tasks. Odds ratios are sample odds ratios of the Experimental relative to the Baseline condition, and p-values are two-sided Fisher’s exact tests.

*Table E. 2. Full self-reported task completion by condition.*

| *Task* | *Baseline (n = 42)* | *Experimental (n = 43)* | *Odds ratio* | *Fisher p* |
|---|---|---|---|---|
| *Task 1: Air quality review* | *90%* | *91%* | *1.03* | *1.00* |
| *Task 2: Fixture and door dimensions* | *69%* | *93%* | *5.98* | *.006* |
| *Task 3: Faulty sensor detection* | *81%* | *93%* | *3.14* | *.117* |

## Acknowledgments

This research was partially supported by the Center for Sand Hazards and Opportunities for Resilience, Energy, and Sustainability (SHORES), funded by Tamkeen under the NYUAD Research Institute Award CG013, and by the NYUAD Center for Interdisciplinary Data Science & AI (CIDSAI), funded by Tamkeen under the NYUAD Research Institute Award CG016.

## Declaration of generative AI and AI-assisted technologies in the manuscript preparation process

During the preparation of this work the authors used ChatGPT in order to improve readability and language. The authors reviewed and edited the output as needed and take full responsibility for the content of the published article.